\pdfoutput=1

\documentclass[10pt,PRE]{article}

\usepackage{changepage}

\usepackage[utf8]{inputenc}

\usepackage{textcomp,marvosym}

\usepackage{fixltx2e}

\usepackage{amsmath,amssymb}

\usepackage{cite}

\usepackage{nameref}
\usepackage{hyperref}

\usepackage{rotating}

\usepackage[toc]{appendix}
\usepackage[aboveskip=1pt,labelfont=bf,labelsep=period,justification=raggedright,singlelinecheck=off]{caption}

\usepackage{subfig}

\date{}

\usepackage{lastpage,fancyhdr,graphicx}
\usepackage{epstopdf}
\usepackage{amsmath}
\usepackage{amsfonts}

\usepackage{color}

\begin{document}
\vspace*{0.35in}

\begin{flushleft}
{\Large
\textbf\newline{Predicting epidemic evolution on contact networks from partial observations}
}
\newline
\\
Jacopo Bindi\textsuperscript{1,*},
Alfredo Braunstein\textsuperscript{1,2,3},
Luca Dall'Asta\textsuperscript{1,2}
\\
\bigskip
\bf{1} Department of Applied Science and Technology, Politecnico di Torino, Corso Duca degli Abruzzi 24, 10129 Torino, Italy
\\
\bf{2} Collegio Carlo Alberto, Via Real Collegio 30, 10024 Moncalieri, Italy
\\
\bf{3} Human Genetics Foundation, Via Nizza 52, 10126 Torino, Italy 
\bigskip

All the authors contributed equally to this work.

*jacopo.bindi@polito.it

\end{flushleft}
\section*{Abstract}
The massive employment of computational models in network epidemiology calls for the development of improved inference methods for epidemic forecast. For simple compartment models, such as the Susceptible-Infected-Recovered model, Belief Propagation was proved to be a reliable and efficient method to identify the origin of an observed epidemics. Here we show that  the same method can be applied to predict the future evolution of an epidemic outbreak from partial observations at the early stage of the dynamics. The results obtained using Belief Propagation are compared with Monte Carlo direct sampling in the case of SIR model on random (regular and power-law) graphs for different observation methods and on an example of real-world contact network. Belief Propagation gives in general a better prediction that direct sampling, although the quality of the prediction depends on the quantity under study (e.g.  marginals of individual states, epidemic size, extinction-time distribution) and on the actual number of observed nodes that are infected before the observation time. 

\section*{Introduction}
Governments and health-care systems maintain costly surveillance programs to report and monitor over time new infection cases for a variety of diseases, from seasonal influenza to the most dreadful viruses such as Ebola. Although surveillance is at the core of modern epidemiology, the early detection of a disease does not automatically guarantee that the fate of the epidemics will be easy to predict, because of the intrinsic stochasticity of the transmission process and the incompleteness of the accessible information. The two issues are often intertwined because a mathematical description that provides a sufficiently accurate prediction at some spatial/temporal scale could become inadequate at another, due to the lack of sufficiently detailed information. For instance, individual-based stochastic compartment models, such as the Susceptible-Infected-Recovered model, are widely used to describe disease transmission in contact networks, but  human interactions have only recently become the object of accurate data mining, and exclusively in small and controlled environments such as schools and hospitals (e.g. by means of the RFID technology) \cite{CLM:CLM12472,salathe2010high,stehlePONE11,mastrandrea2015contact}.
For large-scale epidemic forecast a detailed individual-based description is challenging, therefore researchers and practitioners have resort to coarse-grained metapopulation representations integrated with large-scale datasets on human mobility and real-time estimated parameters \cite{colizza2006modeling,colizza2007predictability,BelikPRX11}. 

Beside the difficulty of obtaining accurate data on human interactions, also the observation of the epidemic progression is usually partial, in particular during the initial stages of an outbreak. For this reason, the ability to use all available information to produce a reliable forecast from early and partial observations is crucial to minimize the impact of a disease, at the same time saving financial resources. Using a simple SIR model, Holme \cite{holme2015time,holme2015information} recently showed that even in the ideal situation in which all information about the structure of the interpersonal network is available, the intrinsic stochasticity of the epidemic process makes prediction of relevant quantities, such as the final outbreak size and the extinction time, very difficult. The quality of the prediction depends on the epidemic parameters (transmission and recovery rates) and on the structure of the underlying network. Predicting the evolution of the epidemics becomes then even more difficult when only partial observation of the state of the individuals is provided. 

In a recent series of works \cite{PhysRevLett.112.118701,1742-5468-2014-10-P10016,Braunstein2016inference}, the problem of inferring the origin of an epidemics in individual-based models from partial observations was investigated. Among the different methods proposed, the Belief Propagation (BP) method \cite{PhysRevLett.112.118701} is not only very reliable and efficient in identifying the origin of an observed epidemics, it also makes possible to easily reconstruct the probability marginals of the individual states at any time, exploiting the causality relations that are generated during the epidemic propagation. Hence, the method can be used to complete the missing information at the time of observation and applied to the problem of epidemic forecasting. 
In the present paper, we will use BP to predict the evolution of a SIR model on a given network from a partial observation of the states of the nodes in the early stage of the dynamics. 

The paper is organized as follows. In Section~\nameref{sec:models} we define direct sampling and Bayesian methods and we introduce the metrics used for validation of the results. Section~\nameref{sec:results} contains a comparison between the prediction obtained using Belief Propagation and Monte Carlo sampling for simulated SIR epidemics on random (regular and power-law) graphs, as well as on a real network of sexual contacts. For these networks, the effectiveness of the methods to predict local (e.g. marginals of individual states) and global (e.g. epidemic size, extinction-time distribution) properties are discussed.  Section~\nameref{sec:methods} reports the description of the main techniques employed in this work, in particular the static factor-graph representation of the epidemic process and the Belief Propagation equations used to evaluate the relevant posterior probabilities of the epidemic process given the observations.


\section*{Inference Models}\label{sec:models}

\subsection*{Prediction from partial observations in the SIR model}

We consider a discrete time susceptible-infected-recovered (SIR) epidemic model on a graph $G=\left( V, E\right)$ that represents a contact network of $N=|V|$ individuals. At each time step of the dynamics a node $i\in V$ can be in one of three possible states: susceptible ($S$), infected ($I$) and recovered ($R$). The state of a node $i$ at time $t$ is represented by a variable $x_i^t\in\left\{S,I,R\right\}$. The stochastic process is defined by a set of parameters $\{\lambda_{ij},\lambda_{ji}\}_{(i,j)\in E}$ and $\{\mu_i\}_{i\in V}$, such that at each time step an infected node $i$ can infect every susceptible node $j$ in his neighborhood $\partial i$ with a probability $\lambda_{ij}$, then recover with probability $\mu_i$ (see Sec.~\nameref{sec:methods} for further details on the dynamics). For a given assignment of the infection parameters and a given initial condition $\mathbf{x}^{0} = \{x_i^0\}_{i\in V}$, a huge number of different realizations of the stochastic process exists, although some of these outcomes are more likely to occur than others. 
Epidemic forecasting consists in providing predictions about how much likely some outcomes are in the form of probability distributions, in particular the probability marginals for the states of individual nodes. In realistic situations, the epidemic forecast is performed at some time after the initial infection event, when a  number of infected cases is discovered in the population. The information available is thus usually localized in time and involves only a fraction of the overall population: we assume that at time $T_{obs}$ the state $x_i^{T_{obs}}$ is made available for a set of nodes $i \in V_{obs} \subset V$ and no information about the state is supplied for the nodes not in $V_{obs}$. In order to focus only on the effects of partial observation, we assume that the structure of the contact network is completely known, and it does not change over time. We remark that, in the case we knew how the network changes over time, we could easily generalize the prediction methods to time-varying networks; unfortunately, the prediction of future contacts in time-varying networks is usually by itself a non-trivial inference problem \cite{cattuto2010dynamics,Holme201297}.

\subsection*{Direct sampling}
Since the SIR stochastic process is Markovian, when $V_{obs}= V$ (complete observation) the probability of the future states ${\mathbf x}^{t}$ for $t > T_{obs}$  can be estimated performing a {\em direct sampling}, that is generating a large number $M_s$ of virtual realizations of the Markov chain from the same initial conditions (a complete observation at $T_{obs}$) and directly estimating the probability of an event from its relative frequency of occurrence in the experiment. In particular, if we call $x_{i,\ell}^t$ the value of the variable $i$ at time $t$ in the $\ell$-th realization of the stochastic process from the same initial conditions, the individual probability marginal $P(x_{i}^{t} = X| {\mathbf x}^{T_{obs}})$ can be estimated from the experimental average 
\begin{equation}
\hat{P}(x_{i}^{t} = X| {\mathbf x}^{T_{obs}}) = \frac{1}{M_s}\sum_{\ell =1}^{M_s} \mathbb{I}\left[ x_{i,\ell}^{t}= X\right]
\end{equation}
that rapidly converges to the correct value with a standard deviation that decreases as $\propto 1/\sqrt{M_s}$.

\begin{figure}[h]
\centering
\includegraphics[width=.9\linewidth]{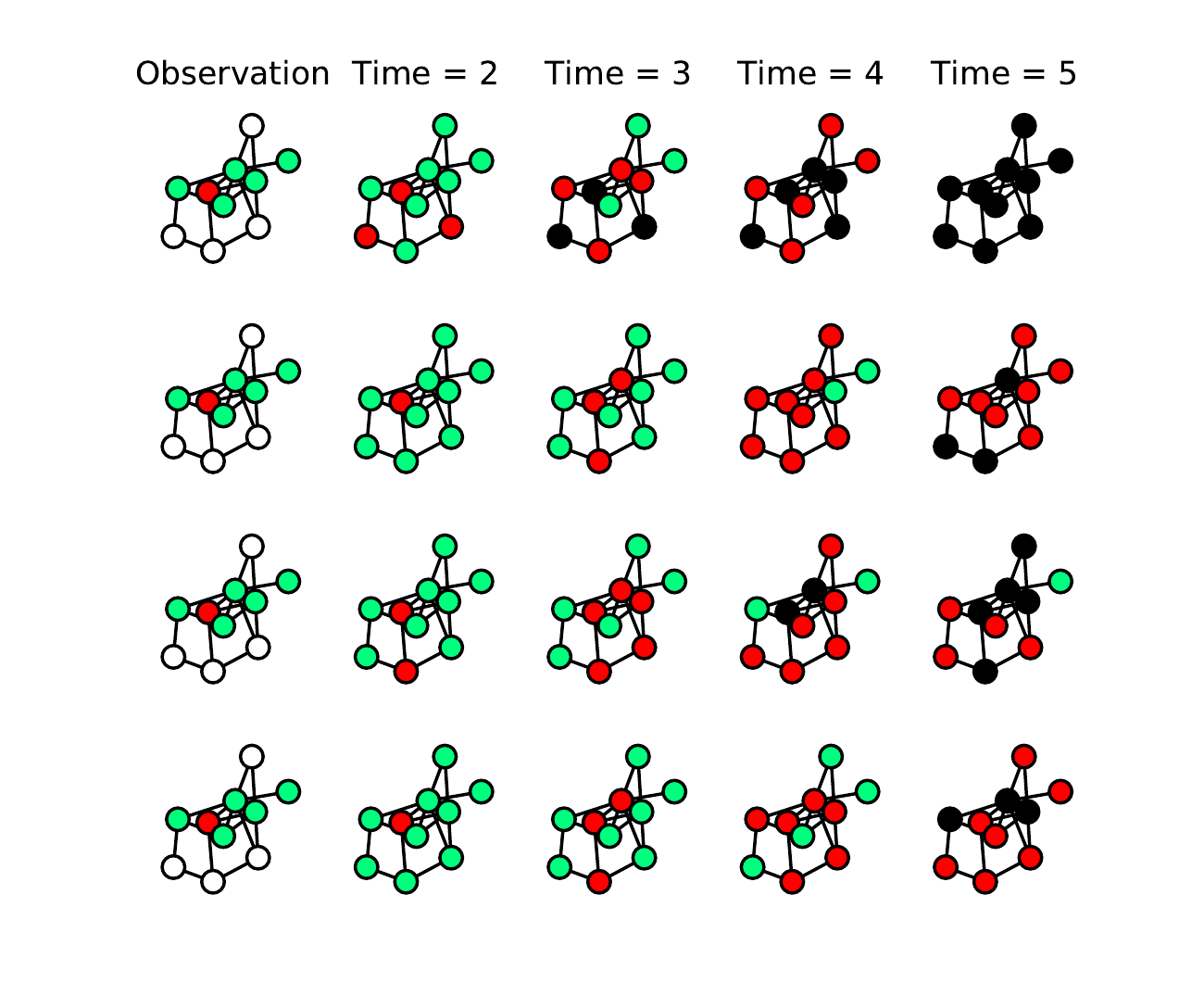}
\caption{Each line represents a different realization for the SIR epidemic process given the (same) incomplete observation of the initial condition.  Configurations in the leftmost column represent the observed state of the system, the other columns represent the time evolution of the epidemic process in that specific realization. Nodes colors: Green = Susceptible, Red = Infected, Black = Recovered, White = Unobserved.}
\label{fig:processExample}
\end{figure}

When ${\mathbf x}^{T_{obs}}$ is only partially known, the uncertainty about the future evolution of an epidemic state is much larger; for instance, Fig.~\ref{fig:processExample} shows five very different evolutions of the epidemic process after the same partial observation. In order to apply the direct sampling method to the case of partial information we first need a way to complete the missing information at $T_{obs}$. In this work we consider two simple ways to choose the states of unobserved nodes at $T_{obs}$:
\begin{itemize}
\item \textbf{random sampling}: given the incomplete observation of the system, the states of unobserved nodes at time $T_{obs}$ are drawn randomly, independently and uniformly with the same probability $1/3$, then direct sampling is performed with such an initial condition. \\  
\item \textbf{density sampling}: given the incomplete observation of the system, the fraction of observed nodes in each state $X\in \{S,I,R\}$ at time $T_{obs}$ is used as an empirical probability to assign, independently and uniformly at random, the state of the unobserved nodes. Direct sampling is then employed to predict future states. The method can be generalized to include dependence on node attributes, such as the degree, by assigning to the unobserved nodes a state with a probability computed from the knowledge of the states of observed nodes with the same attributes. 
\end{itemize}
We remark that unlike the case of complete information, the estimators obtained from these methods through direct sampling have non-zero bias.

\subsection*{A bayesian approach}
The {\em posterior} probability of a configuration $\mathbf{x}^t$ at time $t$ given an observation $\mathbf{x}^{T_{obs}}$ at time $T_{obs}$ can be written as
\begin{equation} 
P\left(\mathbf{x}^{t}|\mathbf{x}^{T_{obs}}\right) = \frac{P\left(\mathbf{x}^t, \mathbf{x}^{T_{obs}}\right)}{P\left(\mathbf{x}^{T_{obs}}\right)} \propto P\left(\mathbf{x}^t, \mathbf{x}^{T_{obs}}|\mathbf{x}^0\right)P\left(\mathbf{x}^0\right),
\label{eq:posterior0}
\end{equation}
where in the last expression we neglected the {\em a priori} probability $P\left({\mathbf x}^{T_{obs}}\right)$ of the observed state that only acts as a normalization constant in our analysis, while $P\left(\mathbf{x}^0\right)$ is the {\em prior} on the initial conditions.
In this Bayesian approach, the prediction of the epidemic evolution after $T_{obs}$ requires to compute the joint probability distribution $P\left(\mathbf{x}^t, \mathbf{x}^{T_{obs}}|\mathbf{x}^0\right)$ of the states at the observation time and at some later time given the initial condition. In principle, this quantity could be evaluated experimentally, by taking into account all possible realizations compatible with the constraints imposed by the dynamics and the observation and discarding the others. However, the number of possible epidemic trajectories of length $t$ grows as $3^{t N}$, making this brute-force approach computationally unfeasible even for small systems and very early observations. An approximate sampling method, that we call {\bf similarity sampling}, is inspired by the Soft-Margin algorithm recently put forward in \cite{antulov2015identification} to infer epidemic origins. The similarity sampling method consists in evaluating $P\left(\mathbf{x}^t, \mathbf{x}^{T_{obs}}|\mathbf{x}^0\right)$ by computing an empirical histogram over a large number of realizations of the epidemic process, each of them contributing with a probability weight that reflects the similarity to the actually observed states at $T_{obs}$. Every node in the set of infected and removed nodes at the time of observation $T_{obs}$ is used as single seed for a given number of realizations. We include unobserved nodes with at least one not susceptible neighbour. We assume to know the initial time of the epidemics within a $\Delta T$ of time steps. Therefore we consider realization with origin in a range $\left[-\Delta T_0, \Delta T_0\right]$. The similarity between a generic realization $\hat{\mathbf{x}}$ and the real one $\mathbf{x}$ is measured by computing the Jaccard similarity function 
\begin{equation}
\phi(\hat{\mathbf{x}}, \mathbf{x}) = \frac{|\mathcal{S}_{I+R}(\hat{\mathbf{x}}^{T_{obs}})\cap \mathcal{S}_{I+R}(\mathbf{x}^{T_{obs}})|}{|\mathcal{S}_{I+R}(\hat{\mathbf{x}}^{T_{obs}})\cup \mathcal{S}_{I+R}(\mathbf{x}^{T_{obs}})|}
\end{equation} 
where $\mathcal{S}_{I+R}(\hat{\mathbf{x}}^{T_{obs}})$ is the set of infected and recovered individuals in realization $\hat{\mathbf{x}}$ at the time of observation. The weight function considered is a gaussian $\propto e^{-(\phi(\hat{\mathbf{x}}, \mathbf{x})-1)^2/a^2}$, where $a$ is a free parameter. Then the individual marginal probability computed by similarity sampling reads:
\begin{equation}
\hat{P}(\hat{x}_{i}^{t} = X| {\mathbf x}^{T_{obs}}) \propto \sum_{\ell =1}^{M_s} \mathbb{I}\left[ \hat{x}_{i,\ell}^{t}= X\right]e^{-(\phi(\hat{\mathbf{x}}, \mathbf{x})-1)^2/a^2}.
\end{equation}

According to \cite{antulov2015identification}, for a fixed value of $a$, we consider a number $M_s$ of realizations such that $\max(|P_{M_s}(x_i)-P_{\frac{M_s}{2}}(x_i)|)<0.1$,  i.e. the maximum of the differences between individual marginals after $M_s$ and $M_s/2$ realizations is smaller than $0.1$. In all results of the present paper we initially set $a=0.125$. If the convergence criterion is not met for $M_s \le 8\times10^5$ we use $a=0.5$. The latter value guarantees the convergence for any instance. Although the method could provide a much more accurate estimate of the individual probability marginals than random and density sampling methods, such an accuracy is usually obtained through fine-tuning of the parameters and requires a very high computational power beyond the aim of this work. 
 
Following the recent work by some of the authors \cite{PhysRevLett.112.118701,1742-5468-2014-10-P10016} we develop here a different approach that consists in addressing the joint probability distribution $P\left(\mathbf{x}^t, \mathbf{x}^{T_{obs}}|\mathbf{x}^0\right)$ as a probabilistic graphical model defined on a static representation of the dynamical trajectories. When the underlying contact network is a tree, the factor graph on which the graphical model is defined can be also reduced to a tree, and the joint probability distribution can be computed exactly by solving a set of local fixed-point equations known as {\bf Belief Propagation} (BP) equations. On more general graphs, the BP equations can be considered as a heuristic algorithm that, under some decorrelation assumptions for the variables, provides a good approximation of the real probability distribution \cite{mezard2009information}. 
The BP equations for the quantity $P\left(\mathbf{x}^{0}|\mathbf{x}^{T_{obs}}\right)$, representing the posterior probability of the initial configuration given an observation at a later time, were derived in Refs.~\cite{PhysRevLett.112.118701,1742-5468-2014-10-P10016}. The BP equations for the more complex graphical model in \eqref{eq:posterior0} are discussed in detail in Sec.~\nameref{sec:methods}. 
We stress that, with the BP approach, the prediction of the future evolution of the epidemics passes through the inference of the (unobserved) dynamical states prior to the time of observation and the reconstruction of the causal relations developed in the dynamics. 

\subsection*{Validation Metrics}

We used the different inference models under study to compute, for every node $i$, the marginal probabilities $P(x_{i}^{t} = X| {\mathbf x}^{T_{obs}})$ with $X\in \{S,I,R\}$. These quantities are then used in the binary classification problem of discriminating whether a node has been infected at a time $t' \leq t$ or not, that turns out to be a relevant measure to quantify the performances of the different prediction methods. In order to do that, we rank the nodes in decreasing order of magnitude of the probabilities $P(x_{i}^{t} = I | {\mathbf x}^{T_{obs}})$ +  $P(x_{i}^{t} = R| {\mathbf x}^{T_{obs}})$ and build a Receiver Operating Characteristic (ROC) curve \cite{PhysRevLett.112.118701,1742-5468-2014-10-P10016}. Starting from the origin of the axes, the ROC curve is obtained from the ordered list of nodes by moving upward by one unit whenever a node is correctly classified as already infected at time $t$ (true positive) or rightward in case it is not (false positive). The {\bf area under the ROC curve} (AUC) expresses the probability that a randomly chosen node that was  infected before time $t$ is actually ranked higher in terms of the corresponding probability marginal than a randomly chosen susceptible one. When the ranking is equal to the real one, the area under the ROC  is $1$, whereas a completely random ordering gives an area equal to $0.5$.

The area under the ROC curve gives indication of the fraction of the correctly classified nodes, but it does not depends much on the actual values of the marginal probabilities. The latter ones have instead a direct effect on a global quantity of crucial important, the size of the epidemic outbreak, i.e. the number of nodes reached by the infection. The {\bf average epidemic size} at time $t$ can be expressed as function of the local marginals as \cite{PhysRevX.4.021024}
\begin{equation}  
size\left(t\right) = \frac{1}{N}\sum_i{\left[P\left(x^t_i=I| {\mathbf x}^{T_{obs}}\right) + P\left(x^t_i=R | {\mathbf x}^{T_{obs}}\right)\right]}.
\end{equation}
The {\bf extinction time distribution} is another relevant global quantity, that cannot be directly computed from the knowledge of the individual probability marginals, and whose characterization on given network structures is a major issue in epidemic studies \cite{holme2013extinction}.  In particular, we are interested in the posterior probability distribution $P\left(T_{ext}| \mathbf{x}^{T_{obs}}\right)$ that the discrete-time epidemic process dies out at time $T_{ext}$ when it is conditioned on the (possibly partial) observation of the state ${\mathbf x}^{T_{obs}}$ at time $T_{obs}$. 
 
A crucial point of the BP algorithm is that it is very convenient for computing local quantities, such as marginal probability laws for the single variables or pair-correlations. Some global quantities, such as the average epidemic size, can be directly computed from the knowledge of the local probability marginals. Interestingly, the quantity $P\left(T_{ext}| \mathbf{x}^{T_{obs}}\right)$ can be expressed as the difference between two terms involving the free energies of the related graphical models when the epidemics are constrained to vanish before time $T_{ext}$ and $T_{ext}-1$, respectively. In Sec.~\nameref{sec:methods} we show that such free energy can be efficiently computed, in the Bethe approximation, as the sum of local terms by means of the BP equations. 
 
 

\section*{Results}\label{sec:results}
\subsection*{Results for individual node classification and epidemic size} 

\begin{figure}[h]
\includegraphics[width=0.9\columnwidth]{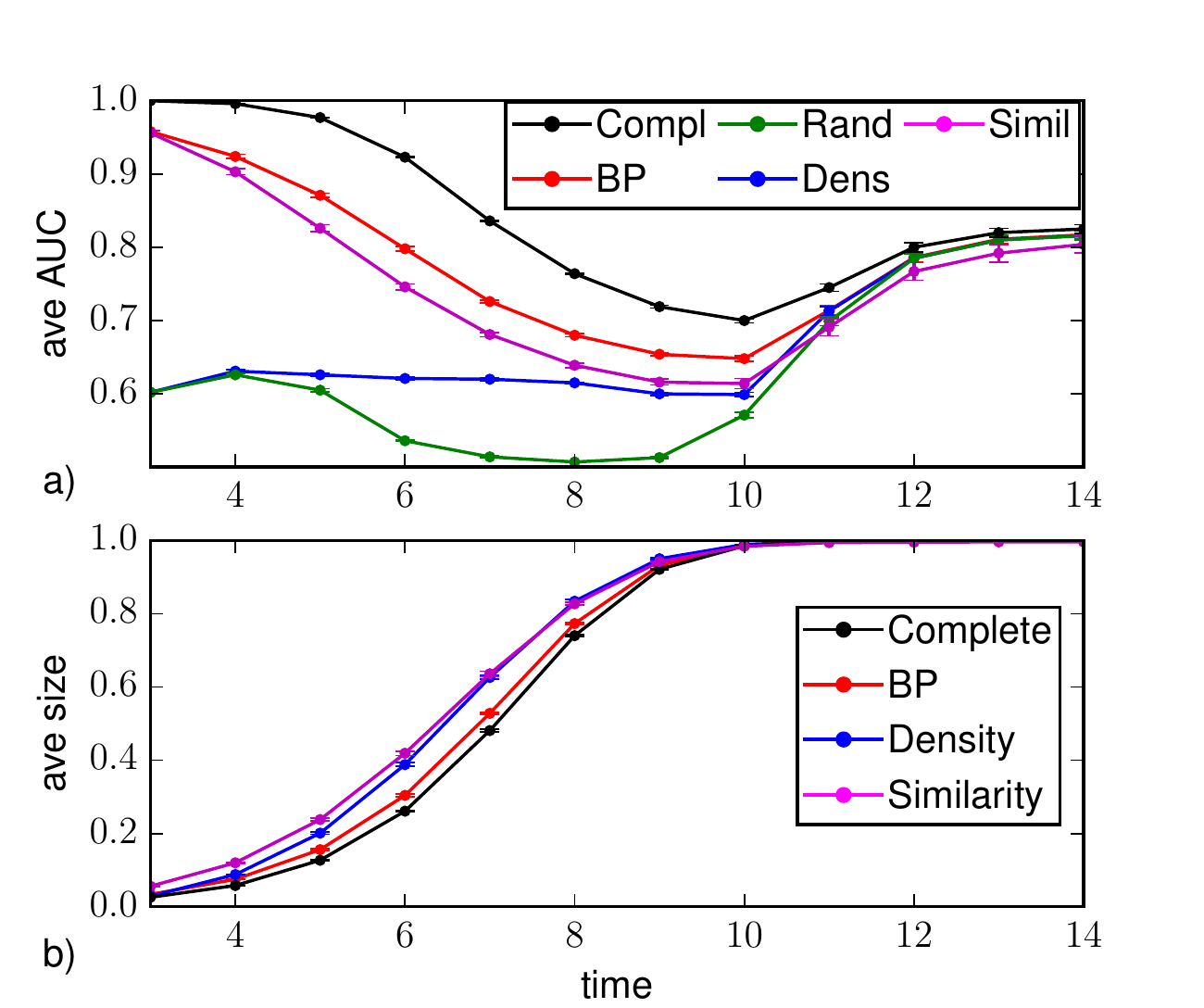}
\caption{ a) Area under the ROC curve as function of the time $t > T_{obs}=3$ on a random regular graph of $N=1000$ nodes and average degree $k=4$. The average is computed over $M_o=10^3$ epidemic realizations (with homogeneous parameters $\lambda=0.7$, $\mu=0.5$); the vertical bars represent the standard error of the mean. The prediction is obtained after the observation at $T_{obs}$ of a $10\%$-fraction of the nodes chosen randomly ({\em random observation}). b) Predicted average epidemic size on random regular graphs ($N=1000$, $k=4$, $\lambda_{ij}=0.7$, $\mu_i=0.5$) as function of time for a random observation of 10\% of the nodes at $T_{obs} = 3$. The inference methods used are direct sampling with complete observation (black), random sampling (green), density sampling (blue), similarity sampling (magenta) and belief propagation (red).}
\label{fig:plot1}
\end{figure}

\paragraph{Random Regular Graphs.} A first set of results for random regular graphs of size $N=1000$ nodes and degree $k=4$ is displayed in Fig.\ref{fig:plot1} and corresponds to the observation of a randomly chosen fraction of $10\%$ of the nodes at $T_{obs}=3$. Fig.\ref{fig:plot1} displays (a) the average values of AUC and (b) the average epidemic size as function of the time steps $t>T_{obs}$ for different prediction methods: random sampling (green), density sampling (blue), similarity sampling (magenta) and Belief Propagation (red). As a reference we also plot results from direct sampling with complete observation (black). The average values are computed on $M_o=10^3$ instances of observations at the same time $T_{obs}$, obtained from independent (and so possibly different) realizations of the epidemic propagations. For each observation, the direct sampling algorithms are performed on $M_s = 2.5\cdot 10^5$ realizations of virtual epidemic processes. We set the error on the initial time $\Delta T_0=1$. The similarity sampling method seldom converges in a number of realizations $M_s=8\cdot10^5$ when $a=0.125$, therefore most of the results are obtained using $a=0.5$. 

Due to the intrinsic stochasticity of the SIR model, node classification is not perfect (AUC values smaller than 1) even in case of direct sampling with complete observation. In fact, the corresponding AUC values rapidly decrease after the observation time and recover only at late times since the epidemics dies out and almost all nodes are either in $R$ or $S$ states. If we interpret the average values of the area under the ROC as a proxy for the epidemic predictability, in the case of a complete observation (best-case scenario) the behavior observed is compatible with the effects due to  epidemic heterogeneity reported in \cite{colizza2006modeling}. 
The most interesting region corresponds to intermediate times, when the predictability of the process is the lowest. Fig.\ref{fig:plot1}a shows that the Belief Propagation technique with partial observation gives values of averaged AUC that are closer to those from complete observation than the other methods. BP and similarity sampling perform largely better in the first stage after the observation, corresponding to the exponential outbreak phase\cite{pastor2015epidemic}. In particular similarity sampling gives an AUC value similar to BP at the time of observation, but a lower AUC value in the subsequent time steps.  

Fig.~\ref{fig:plot1}b shows that density sampling strongly overestimates the average epidemic size with respect to results from complete observation; this is probably an effect of the homogeneous deployment over the graph of infected nodes used to complete the information, that favors a larger epidemic spreading. Disregarding existing correlations between the 90\% of the nodes, this scheme could lead to the overestimation of the probability of being infected -- in a similar way to mean field approximations. In similarity sampling the overestimation of the epidemic size is due both to this procedure to set the seeds of the $M_s$ virtual epidemic realizations and to the approximation on the initial time. Belief Propagation also slightly overestimates the epidemic size, but we think this is essentially due to the fact that in most of the instances the algorithm does not properly converge to the correct marginals.

\begin{figure}[h]
\centering
\includegraphics[width=0.9\columnwidth]{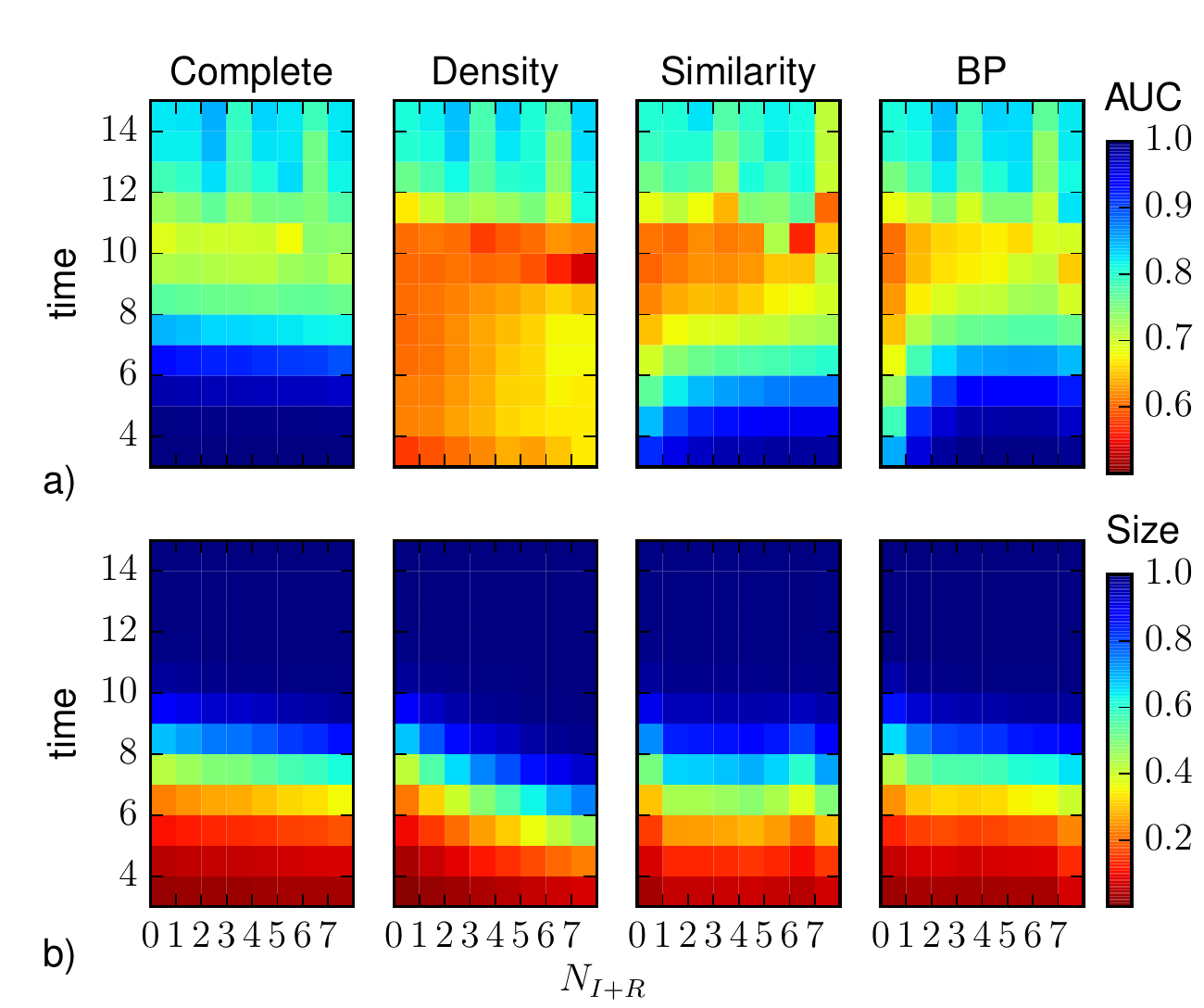}
\caption{a) The heatplots represent the average AUC as function of time and of the number of observed nodes that were infected before $T_{obs}$, computed by density sampling, similarity sampling, belief propagation. b) The average epidemic size predicted by density sampling, similarity sampling and belief propagation is also shown as function of the number of infected and recovered nodes in the observation. As a reference, in both panels, we plot results obtained, for the same realizations of the SIR process, by direct sampling with complete observation. The horizontal axis refers to the number of infected or recovered nodes present in the $10\%$ observation (also in the case of complete observation). }
\label{fig:heat1}
\end{figure}

The heatplots in Fig.\ref{fig:heat1} display the same set of data classified as function of the number of observed nodes that were infected before the observation time, respectively for density sampling, similarity sampling and belief propagation. Results for direct sampling with complete observation are presented as a reference. In the case of the average AUC (Fig.\ref{fig:heat1}), BP performs better than both density sampling and similarity sampling in all regimes, in particular the performance is very good in the first steps after the observation, almost independently of the actual number of infected and recovered nodes in the observation. For all methods the results slightly improve when a larger number of nodes reached by the epidemics is observed at $T_{obs}$. For the average epidemic size, Fig.~\ref{fig:heat1}b shows that the early-stage prediction by density sampling is negatively affected by the observation of a larger number of infected and recovered nodes. The opposite occurs, though to a lesser extent, for BP: when few infected nodes are observed BP overestimates the epidemic size, the worst prediction by BP giving an average size about 20\% larger than that obtained from complete observation. The deviation observed by similarity sampling is also more evident when a lower number of infected and recovered nodes is observed, but the overestimation is more homogeneously distributed. Interestingly, the poor performance at large times is localized on realizations in which only a few of the observed nodes already got infected at $T_{obs}$. 

\paragraph{Barab\'{a}si-Albert random graph.}
In the case of heterogeneous graphs, such as those obtained with the Barab\'{a}si-Albert (BA) growing network model, in addiction to the {\em random observation}, it is interesting to define other observation schemes for the same density of observed nodes: 
\begin{itemize}
\item {\em degree-based observation}: nodes are observed in descending order of their degree; 
\item {\em local observation}: a connected cluster of observed nodes is generated from a randomly chosen infected node.
\end{itemize}
We investigated the effect of different observation schemes on random sampling, density sampling, similarity sampling and BP. 

\begin{figure}[ht]
\centering
\includegraphics[width=0.9\columnwidth]{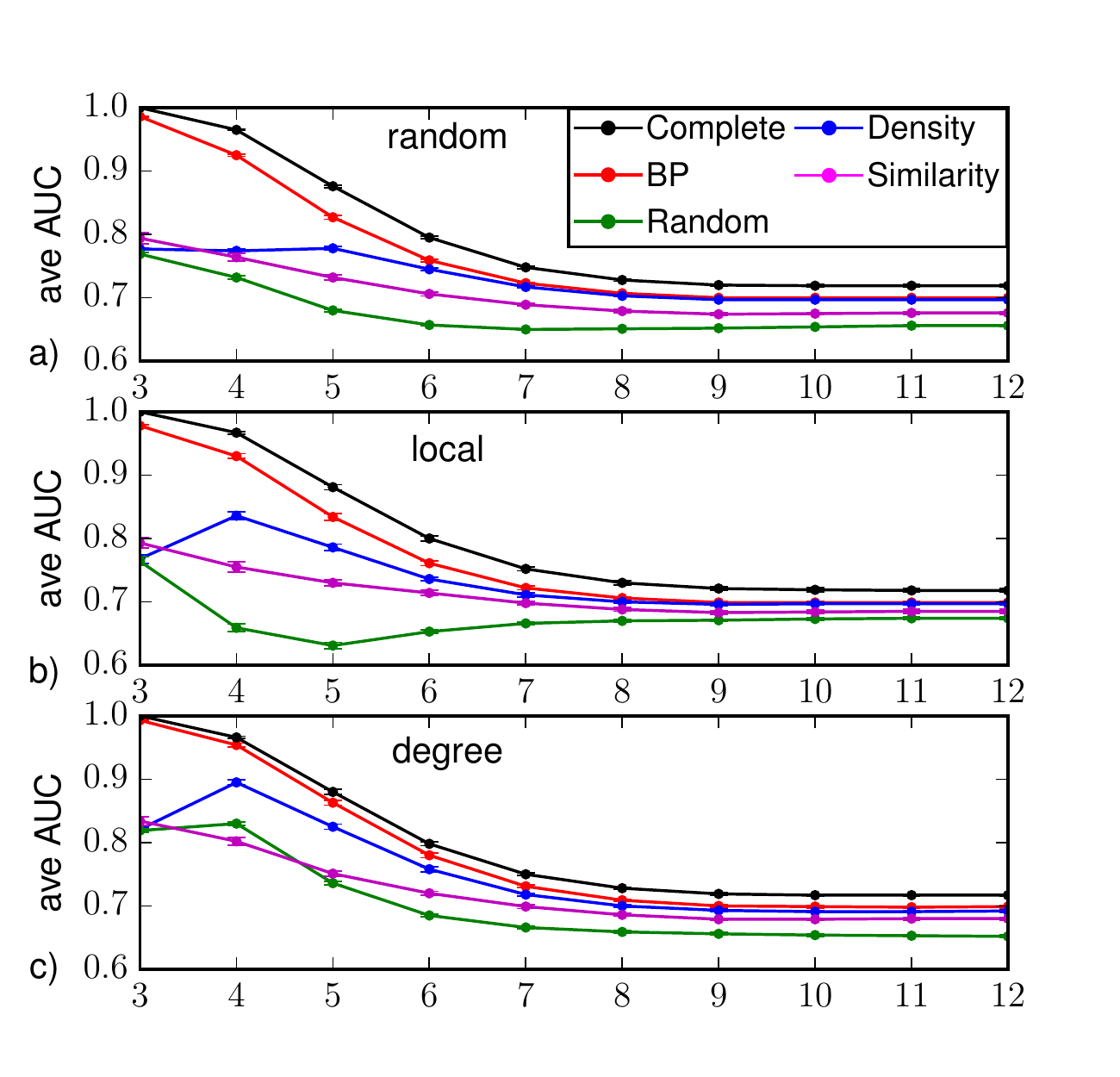}
\caption{Area under the ROC curve as function of the time $t > T_{obs}=3$ on a Barab\'{a}si-Albert random graph of $N=1000$ nodes and average degree $\left\langle k\right\rangle\approx 4$ (with homogeneous epidemic parameters $\lambda=0.5$, $\mu=0.6$), in the case of observation of a $30\%$-fraction of (a) nodes chosen at random uniformly and independently, (b) nodes forming a connected subgraph, (c) the most connected nodes.  The average is computed over $M=201$ epidemic realizations. The inference methods used are direct sampling with complete observation (black), random sampling (green), density sampling (blue), similarity sampling (magenta) and belief propagation (red).}
\label{fig:percBA}
\end{figure}

\begin{figure}[ht]
\centering
\includegraphics[width=0.9\columnwidth]{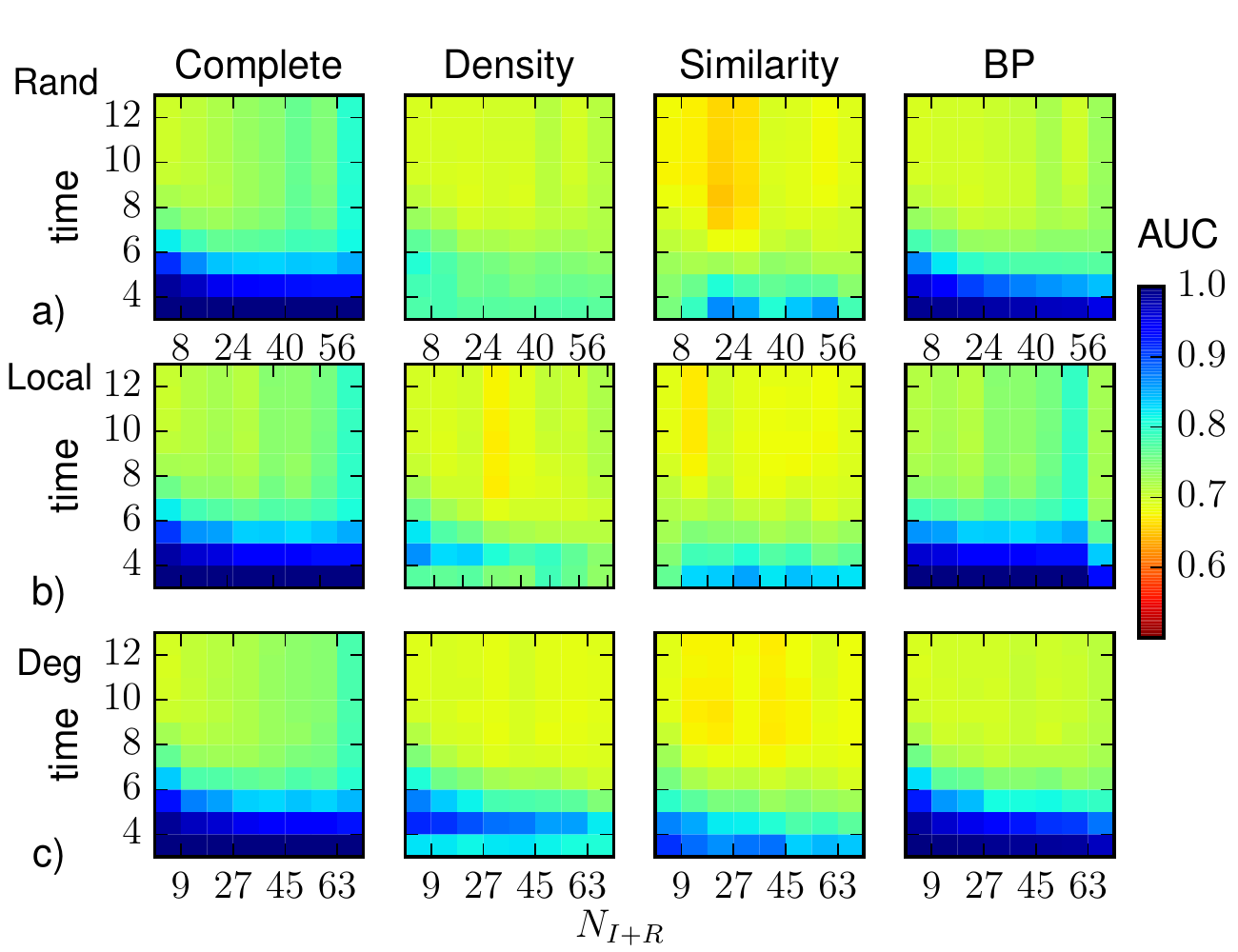}
\caption{ The heatplots represent the average AUC as function of time and of the number of observed nodes that were infected before $T_{obs}=3$, computed by density sampling, similarity sampling, belief propagation, on a Barab\'{a}si-Albert random graph of $N=1000$ nodes and average degree $\left\langle k\right\rangle\approx 4$ with homogeneous parameters $\lambda=0.5$, $\mu=0.6$. As a reference, we also plot results obtained, for the same realizations of the SIR process, by direct sampling with complete observation.
The prediction is obtained after the observation at $T_{obs}$ of a $30\%$-fraction of (a) nodes chosen at random uniformly and independently, (b) nodes forming a connected subgraph, (c) the most connected nodes. The horizontal axis refers to the number of infected or recovered nodes present in the $30\%$ observation (also in the case of complete observation).
} 
\label{fig:heatBA}
\end{figure}

The results  for the average AUC, obtained with observation of $30\%$ of the nodes at $T_{obs}=3$, are reported in Figs.~\ref{fig:percBA}-\ref{fig:heatBA}. In the case of complete observation, direct sampling produces monotonically decreasing AUC values for increasing times. The reason is that in finite size networks the parameters chosen give a non-zero probability of finding susceptible nodes in the last stage of the epidemic evolution, then wrong predictions are possible and the AUC remains considerably below one. For random observation, Fig.\ref{fig:percBA}a shows that Belief Propagation always gives larger AUC values than the other sampling methods, especially in the first stage of the epidemics, i.e. during the exponential outbreak. The same behavior is found plotting the results as function of the actual number of observed nodes (see heatplots in Fig.\ref{fig:heatBA}a) that were already infected at the time of observation; in particular, the performances of BP are better when this number is small. 

Fig.~\ref{fig:percBA}b-\ref{fig:heatBA}b report results obtained with the observation of a $30\%$-fraction of nodes forming a connected subgraph. The overall results are very similar to those with random observation, even though density sampling and BP perform slightly better in the time steps immediately after the observation, while similarity sampling is slightly worse in the same regime.  
A degree-based observation is particularly convenient for heterogenous networks. Fig.\ref{fig:percBA}c shows that the average values of the ROC area increase in the first stage of the epidemics for all prediction methods, in particular the difference between values obtained by BP and those from direct sampling with complete observation is less than $2\%$. The results reported in the heatplots (see Fig.\ref{fig:heatBA}c) are qualitatively similar to those from the other observation schemes, with slightly better prediction performances at early times when the number of infected nodes in the observation is small. This is possibly due to the fact that these cases correspond to smaller epidemics whose initial evolution is more predictable. \\

\begin{figure}[h]
\centering
\includegraphics[width=0.9\columnwidth]{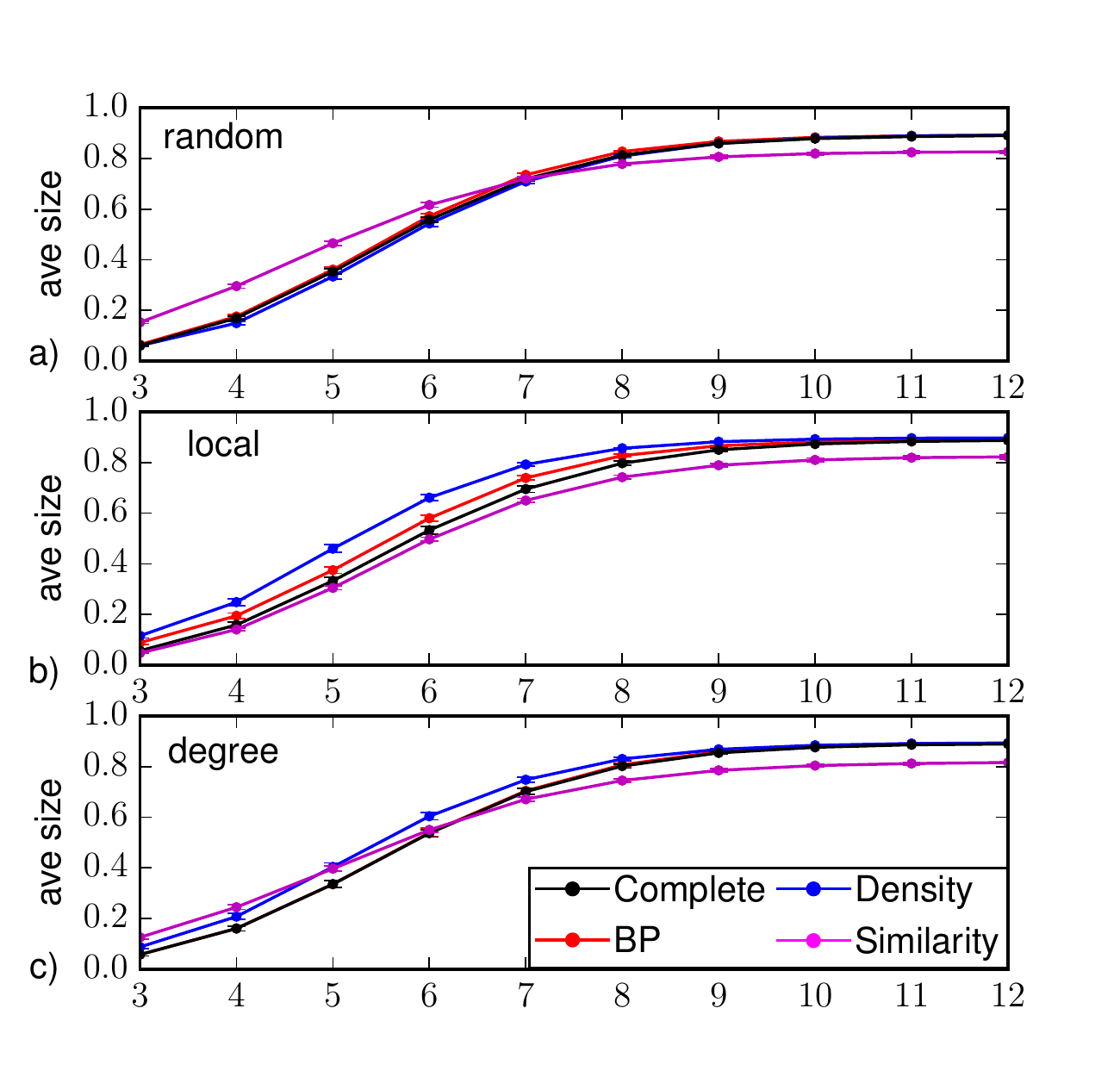}
\caption{Predicted average epidemic size as function of the time $t > T_{obs}=3$ on a Barab\'{a}si-Albert random graph of $N=1000$ nodes and average degree $\left\langle k\right\rangle\approx 4$ (with homogeneous epidemic parameters $\lambda=0.5$, $\mu=0.6$), in the case of observation of a $30\%$-fraction of (a) nodes chosen at random uniformly and independently, (b) nodes forming a connected subgraph, (c) the most connected nodes. The average is computed over $M=201$ epidemic realizations. The inference methods used are direct sampling with complete observation (black), density sampling (blue), similarity sampling (magenta) and belief propagation (red).
} 
\label{fig:sizeBA}
\end{figure}
\begin{figure}[h]
\centering
\includegraphics[width=0.9\columnwidth]{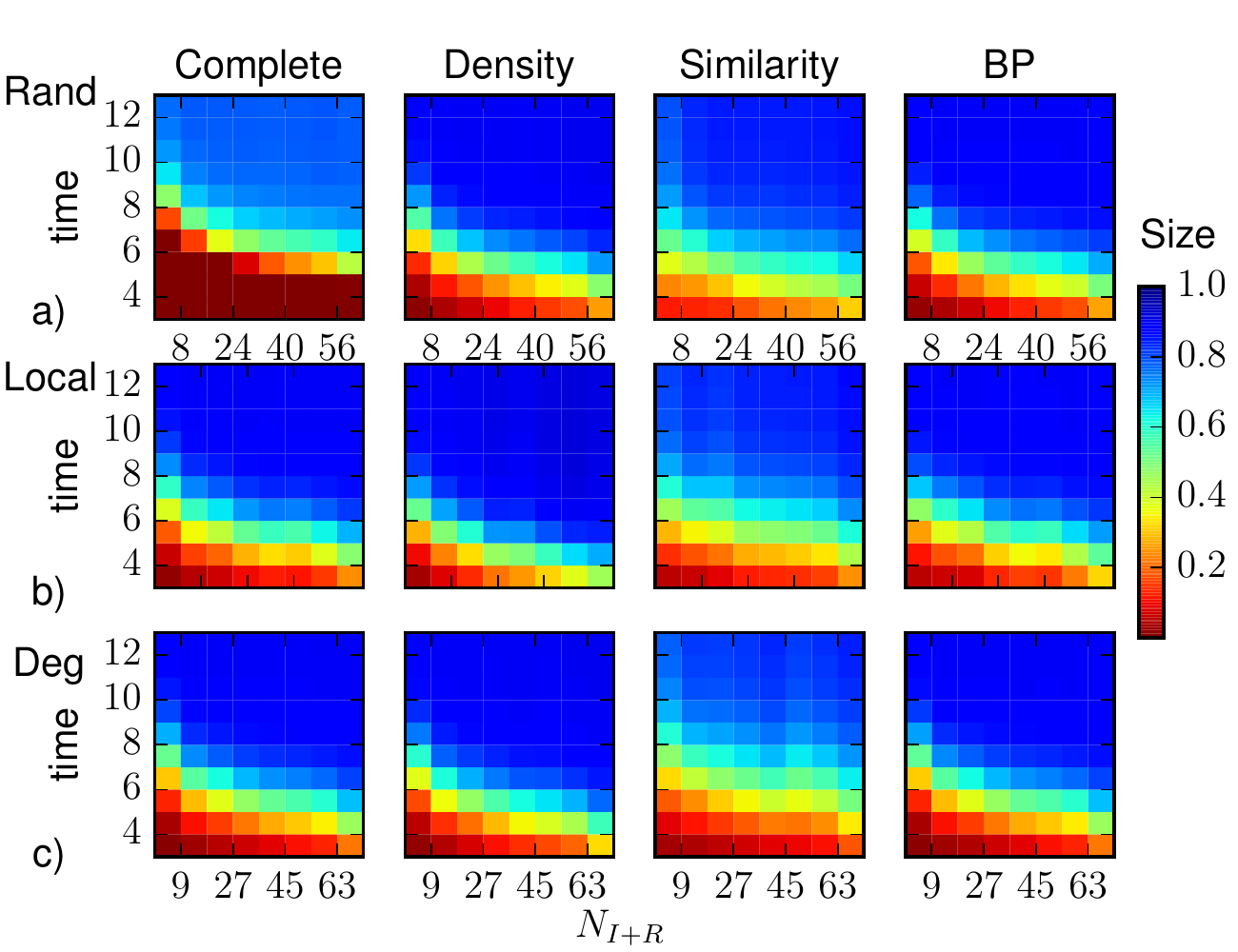}
\caption{
The heatplots represent the average epidemic size as function of time and of the number of observed nodes that were infected before $T_{obs}=3$, computed by density sampling, similarity sampling, belief propagation, on a Barab\'{a}si-Albert random graph of $N=1000$ nodes and average degree $\left\langle k\right\rangle\approx 4$ with homogeneous parameters $\lambda=0.5$, $\mu=0.6$. As a reference, we also plot results obtained, for the same realizations of the SIR process, by direct sampling with complete observation.
The prediction is obtained after the observation at $T_{obs}$ of a $30\%$-fraction of (a) nodes chosen at random uniformly and independently, (b) nodes forming a connected subgraph, (c) the most connected nodes. The horizontal axis refers to the number of infected or recovered nodes present in the $30\%$ observation (also in the case of complete observation).
}
\label{fig:HeatsizeBA}
\end{figure}

Results on the prediction of the average epidemic size on Barab\'{a}si-Albert networks are reported in Fig.~\ref{fig:sizeBA} and Fig.\ref{fig:HeatsizeBA}, except for random sampling that strongly overestimates the size in all regimes and observation schemes. With a random observation scheme (see Fig.~\ref{fig:sizeBA}a), density sampling and BP provide very accurate prediction along the whole dynamics, while similarity sampling provides strong overestimate of the size value at early time and underestimate at late times. Fig.\ref{fig:HeatsizeBA}a suggests that for both density sampling and BP, the accuracy is lower when a small number of infected and recovered nodes is observed. When the number of nodes reached by the infection at $T_{obs}$ is larger, BP performs better than density sampling ($4.5\%$ of the nodes larger than the direct sampling with complete observation). The very bad results of similarity sampling at late times are mostly due to a very strong underestimation of the average size when the observation contains only few infected/recovered nodes. On the contrary at early times overestimate appears when a large number of infected/recovered nodes is observed.

In Fig.\ref{fig:sizeBA}b we show the prediction of the average epidemic size when the partial observation is performed considering a connected subgraph of 30\% of the nodes. In this case all methods overestimate the epidemic size, with BP performing considerably better than the others. The poor performances of density sampling are expected because it completely neglects the topological information in the observation. For example if infected nodes are surrounded by susceptible ones, the probability of infection for unobserved nodes is lower, but this is not taken into account in the density sampling approach. BP performs instead poorly when there are very few infected nodes in the observed area. This is expected, because in such a situation this method is not able to correctly reconstruct missing information. Finally, similarity sampling gives good results for small and intermediate time steps but again it strongly deviates at large times, mostly because of observations with few infected/recovered nodes (see Fig.\ref{fig:HeatsizeBA}b).

We already noticed that Belief Propagation performs very well in the case of a degree-based observation; this is true also for the epidemic size prediction, as shown in Fig.\ref{fig:sizeBA}c. The difference between the average epidemic size predicted by Belief Propagation and the one obtained by direct sampling with complete observation is less than 2\% of the nodes in the network. Instead, density sampling overestimates the average epidemic size, especially in the first epidemic outbreak and for a large number of infected and recovered nodes in the observation (see Fig.\ref{fig:HeatsizeBA}c). Density sampling does not make use of the connectivity knowledge, which is a valuable information: an observed highly connected node is more likely to be infected, ignoring this fact leads to assign the same infection probability of the hubs to every node in the network, leading to larger predicted epidemic sizes. In this respect, one could expect that better results could be obtained simply by introducing a degree-dependence in the infection probability inferred from the observation; nevertheless, preliminary results show no significant improvement in the quality of the prediction. 

\subsection*{Results for the extinction time distribution}
\begin{figure}[tbh]
\includegraphics[width=0.9\linewidth]{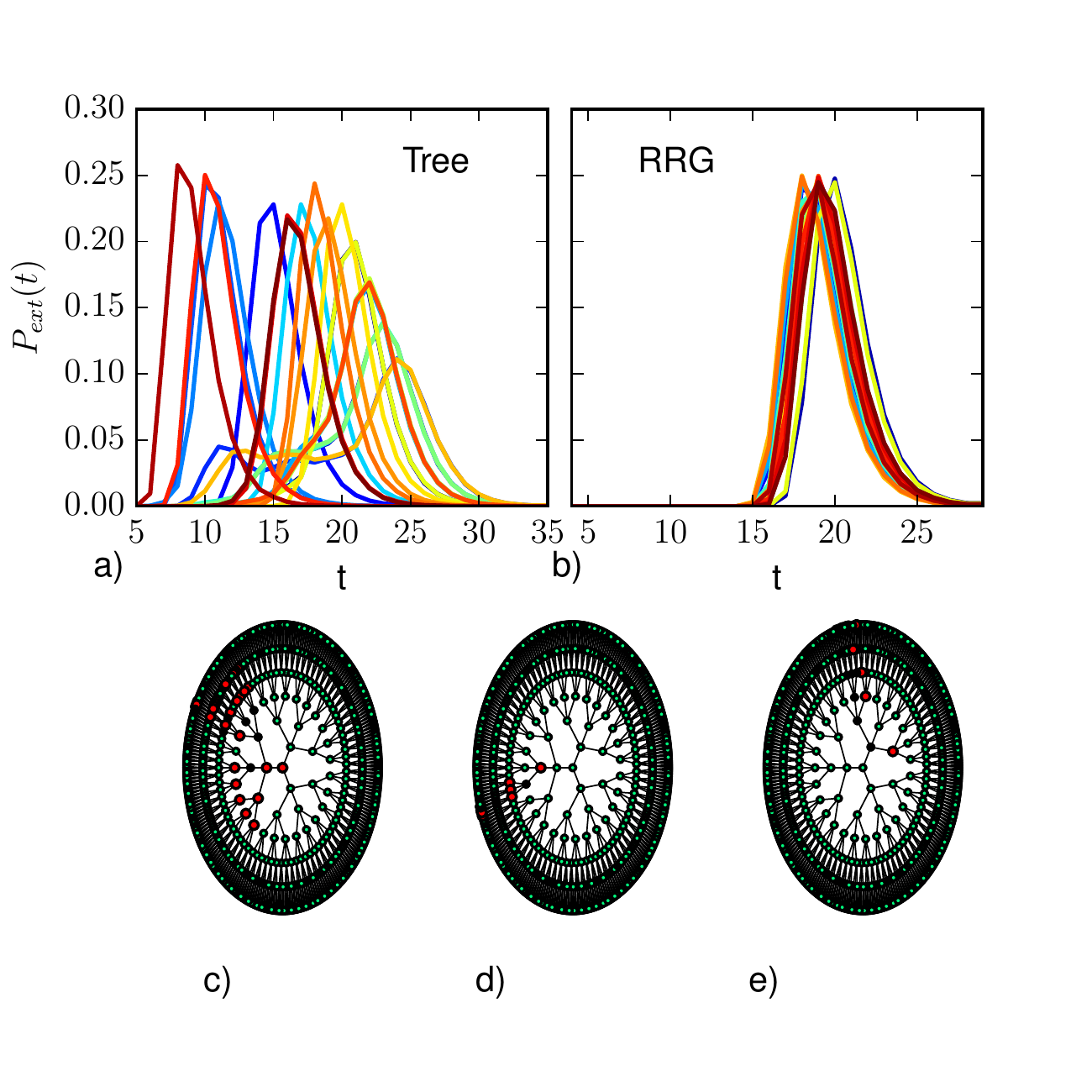}
\caption{The extinction time distributions for different complete observations: a) on trees with branching ratio $k=3$ and $N=1092$ (epidemic parameters $\lambda =0.7$, $\mu =0.5$, and observation time $T_{obs}=5$); b) on random regular graphs of degree $4$ and $N=1000$ nodes (epidemic parameters $\lambda =0.7$, $\mu =0.5$ and observation time $T_{obs}=4$). Panels (c)-(e) illustrate similar realizations of the epidemic process at $T_{obs}$ on a tree graph corresponding to rather different predicted extinction time distributions with maximum value respectively at $T=21$ (c), $T=23$ (d), and $T=16$ (e). Nodes color: Green= Susceptible, Red= Infected, Black= Recovered.}
\label{fig:Ext}
\end{figure}

The extinction time distribution is a global feature of the epidemic process, that can strongly depend on the epidemic parameters, the initial conditions and the topological structure of the underlying contact network. Here we are interested in predicting the probability distribution for the extinction time when a (possibly partial) observation is provided. Even in the case of complete observation, the results are highly non-trivial, in particular on networks with peculiar topological structure. Fig.~\ref{fig:Ext} shows the extinction time distribution $P_{ext}(t) = P\left(t=T_{ext}| \mathbf{x}^{T_{obs}}\right)$ for regular trees (a) and regular random graphs (b). In the case of trees the probability distribution is highly variable: depending on the observation,  the width and the maximum value of the distribution can change significantly. Figs.~\ref{fig:Ext}c-\ref{fig:Ext}e show three different realizations at the time of observation $T_{obs}$. In terms of the number of infected node and their average degree, the snapshots in panels (c) and (e) are similar, but their extinction time probability distributions are rather different ($T_{peak}=16$ and $T_{peak}=23$). On the contrary, despite the very different realizations at $T_{obs}$, snapshots in panels (c) and (d) correspond to similar distributions ($T_{peak}=23$ and $T_{peak}=21$). This is due to the arrangement of infected and recovered nodes at the time of observation: the configuration in Fig.\ref{fig:Ext}e does not allow to access the root of the tree, so the epidemics is limited and diffusion to other branches of the graph is blocked. In Figs.\ref{fig:Ext}c-\ref{fig:Ext}d, instead, the epidemics can spread throughout the graph, causing the  distribution to reach a maximum at larger times.
The heterogeneity of the extinction time distribution is peculiar of trees and graphs with topological bottlenecks, while random graphs, or graphs with small-world properties in general, are characterized by very similar distributions for different realizations of the epidemic process (with same epidemic parameters and observation).

\begin{figure}[tbh]
\includegraphics[width=0.9\linewidth]{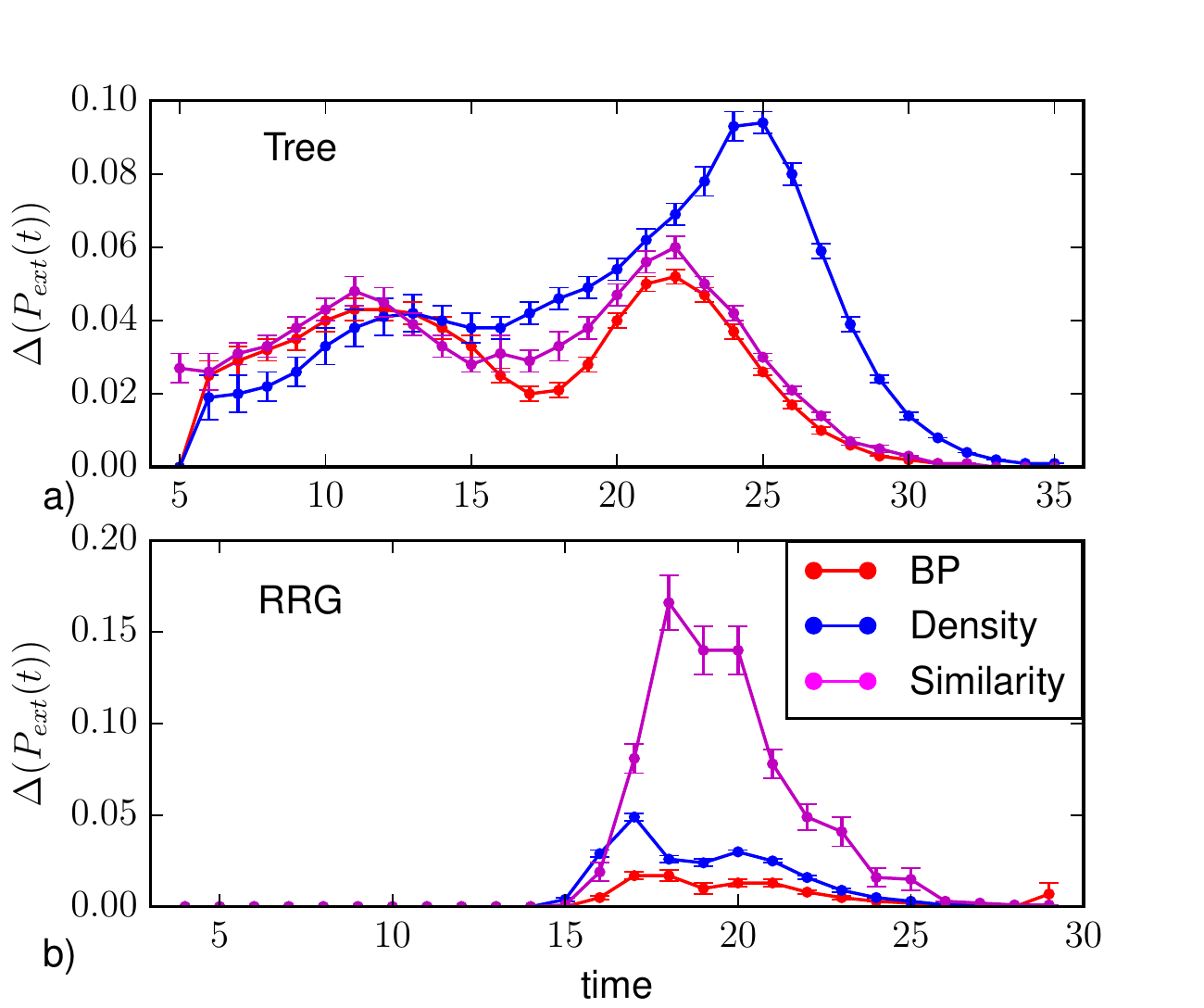}
\caption{Absolute value of the difference between the extinction time distribution $P_{ext}(t)$ computed from direct sampling with complete information and those calculated with density sampling (blue), BP (red) and similarity sampling (magenta). a) On trees of $N=1092$ nodes, with branching ratio $3$ ($\langle k\rangle \approx 2$) and with uniform epidemic parameters $\lambda =0.7$, $\mu =0.5$. The partial observation is performed sampling uniformly the state of $10\%$ of the nodes at $T_{obs}=5$ and averaging over $M_o=210$ such realizations.  b) On random regular graphs of $N=1000$ nodes and degree $k =4$ with uniform epidemic parameters $\lambda =0.7$, $\mu =0.5$. The partial observation is performed sampling uniformly the state of $30\%$ of the nodes at $T_{obs}=4$ and averaging over $M_o=150$ such realizations.}
\label{fig:Ext1}
\end{figure}

\begin{figure}[tbh]
\includegraphics[width=0.9\linewidth]{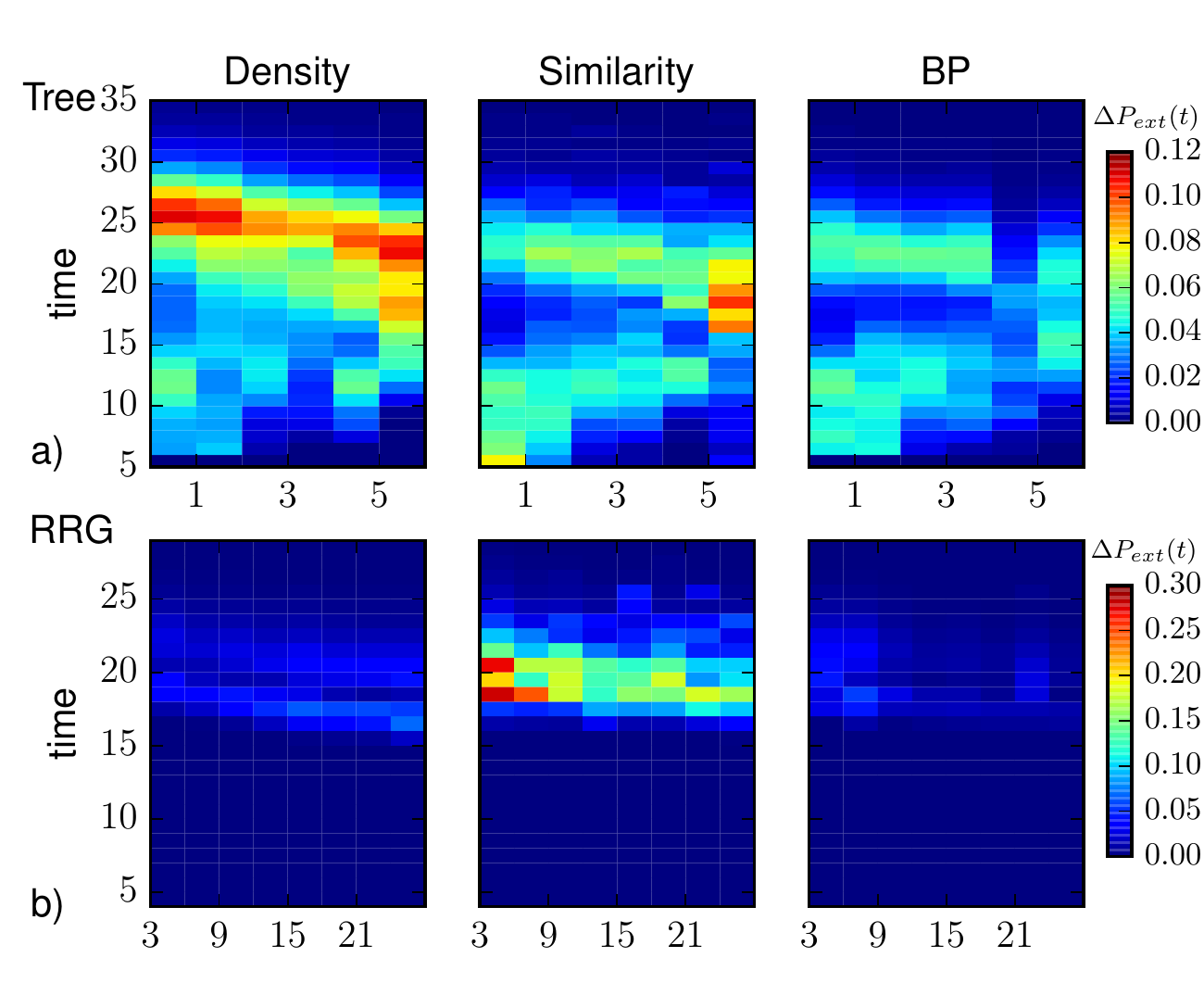}
\caption{Absolute value of the difference between the extinction time probability distribution $P_{ext}(t)$ computed from direct sampling with complete information and those calculated with density sampling, BP and similarity sampling as a function of the number of infected and recovered nodes in the observed subset of nodes. 
a) On trees of $N=1092$ nodes, with branching ratio $3$ ($\langle k\rangle \approx 2$) and with uniform epidemic parameters $\lambda =0.7$, $\mu =0.5$. The partial observation is performed sampling uniformly the state of $10\%$ of the nodes at $T_{obs}=5$ and averaging over $M_o=210$ such realizations.  b) On random regular graphs of $N=1000$ nodes and degree $k =4$ and with uniform epidemic parameters $\lambda =0.7$, $\mu =0.5$. The partial observation is performed sampling uniformly the state of $30\%$ of the nodes at $T_{obs}=4$ and averaging over $M_o=150$ such realizations.
}
\label{fig:Ext2}
\end{figure}

The results on the prediction of the extinction time distribution from partial observations is shown in Fig.~\ref{fig:Ext1}. Motivated by the observed strong variability of the extinction time distribution, we first considered the case of regular trees of branching ratio equal to $4$ (average degree $\langle k\rangle \approx 2$). The partial observation was obtained sampling randomly the state of $10\%$ of the nodes at $T_{obs} = 5$. Fig~\ref{fig:Ext1}a displays the average difference between the extinction time distribution predicted using direct sampling with complete observation and that obtained using Belief Propagation (red), density sampling (blue), and similarity sampling (magenta). All methods present two regions of higher discrepancy with respect to the prediction with complete observation. As shown by the heatplots in Fig.\ref{fig:Ext2}b, this is usually due to an underestimation of the probability of extinction in the early stage of propagation and to an overestimation of the probability of extinction at large times. BP is usually able to qualitatively identify the most probable extinction time even when the other methods instead assign more probability mass to much larger times. Heatplots show that the two-peak discrepancy is especially due to observations with few infected and recovered nodes, while the discrepancies between the distributions move mostly at intermediate times when this number is increased. BP performs better than the other methods at every time step, although it presents the same qualitative weaknesses. Interestingly, the similarity sampling method overestimates the probability for the epidemics to die out at early time step. In fact, a fraction of the epidemics with a large similarity index to the observed incomplete snapshot immediately dies out after $T_{obs}$, leading to an overestimation of the extinction probability at early time steps. 

Fig.~\ref{fig:Ext1}b and Fig.~\ref{fig:Ext2}b display the same analysis in the case of random regular graphs of degree $k=4$ with partial observation of the $30\%$ of the nodes at $T_{obs} = 4$. Although all prediction methods under study are able to reproduce the existence of a unique peak, there are remarkable quantitative differences with the results from direct sampling with complete observation. The BP algorithm provides the best performances, in particular for observations with a large number of infected and recovered nodes. For a low number of infected and recovered nodes, instead, BP gives a larger average difference with respect to density sampling. This effect is mostly due to the non-convergence of the BP algorithm in some instances of the epidemic process, leading an overestimation of the probability of long extinction times. At the time steps close to the peak the similarity sampling give a larger average difference with respect to density sampling and BP. The similarity sampling gives the largest average difference. The main contribution to the average difference comes when low number of infected and recovered nodes are observed. In this case information provided by the observation is insufficient for similarity sampling.

\subsection*{A case study of real contact network}
We consider a real network dataset of the sexual encounters of internet-mediated prostitution \cite{rocha2010information,rocha2011simulated}, that was obtained analyzing a Brazilian web community exchanging information between male sex buyers. The original dataset is in the form of a bipartite temporal network, in which an edge between a ``sex buyer'' A and ``sex seller'' B is drawn if A posted a comment in a thread about B. The dataset covers the period September 2002 to October 2008 (2,232 days) and 50,185 contacts are recorded between 6,642 sex sellers and 10,106 sex buyers. In our analysis, we do not consider separate classes of vertices and we focus on a sample network comprising a time window between day 1000 and day 1100. The resulting network (SC) has $N=1293$ nodes, $E=1571$ edges, average degree $\left\langle k\right\rangle \approx 2.4$ and maximum degree $k_{max}=55$.

\begin{figure}[tbh]
\centering
\includegraphics[width=0.9\columnwidth]{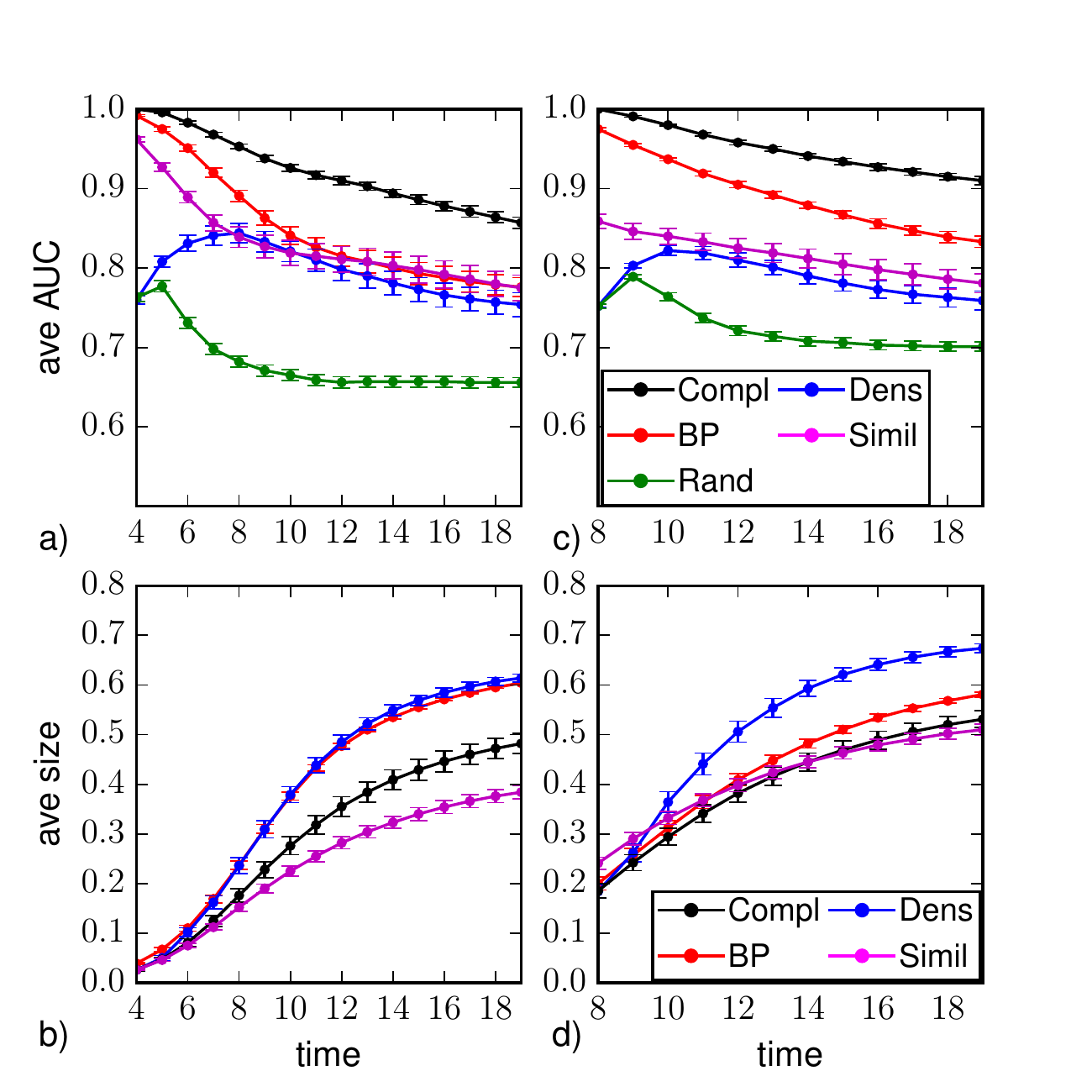}
\caption{
Average area under the ROC curve (a,c) and average epidemic size (b,d) as function to the time $t \ge T_{obs}$ for SIR dynamics ($\lambda=0.5$, $\mu=0.4$) on the SC network. Results are obtained with random sampling (green), density sampling (blue), similarity sampling (magenta) and Belief Propagation (red) from a random observation of $30\%$ of the nodes at $T_{obs}=4$ (a,b) and $T_{obs}=8$ (c,d). In all plots direct sampling from a complete observation is shown for comparison (black).}
\label{fig:casetudy1}
\end{figure}

We study the predictability of the epidemic evolution on a static projection of the sexual contact network when the observation takes place at times $T_{obs}=4,8$ as representatives of early and later time observation. In both cases, density sampling and random sampling make unreliable predictions of the classification of individual states of the nodes (see Fig.\ref{fig:casetudy1}a-\ref{fig:casetudy1}c). For $T_{obs}=4$, BP gives good results only in the time steps immediately after the observation, then the performances rapidly deteriorate. BP results slightly improve increasing the observation time. Nevertheless BP is better than other methods. For the average epidemic size, Fig.~\ref{fig:casetudy1}b shows that similarity sampling gives the best prediction at $T_{obs}=4$ (though underestimating the epidemic size), whereas BP performs as bad as density sampling (and random sampling even worse). BP results improve considerably for $T_{obs}=8$ while similarity sampling turns out to overestimate the epidemic size at early times (Fig.~\ref{fig:casetudy1}d).

\begin{figure}[tbh]
\centering
\includegraphics[width=0.9\columnwidth]{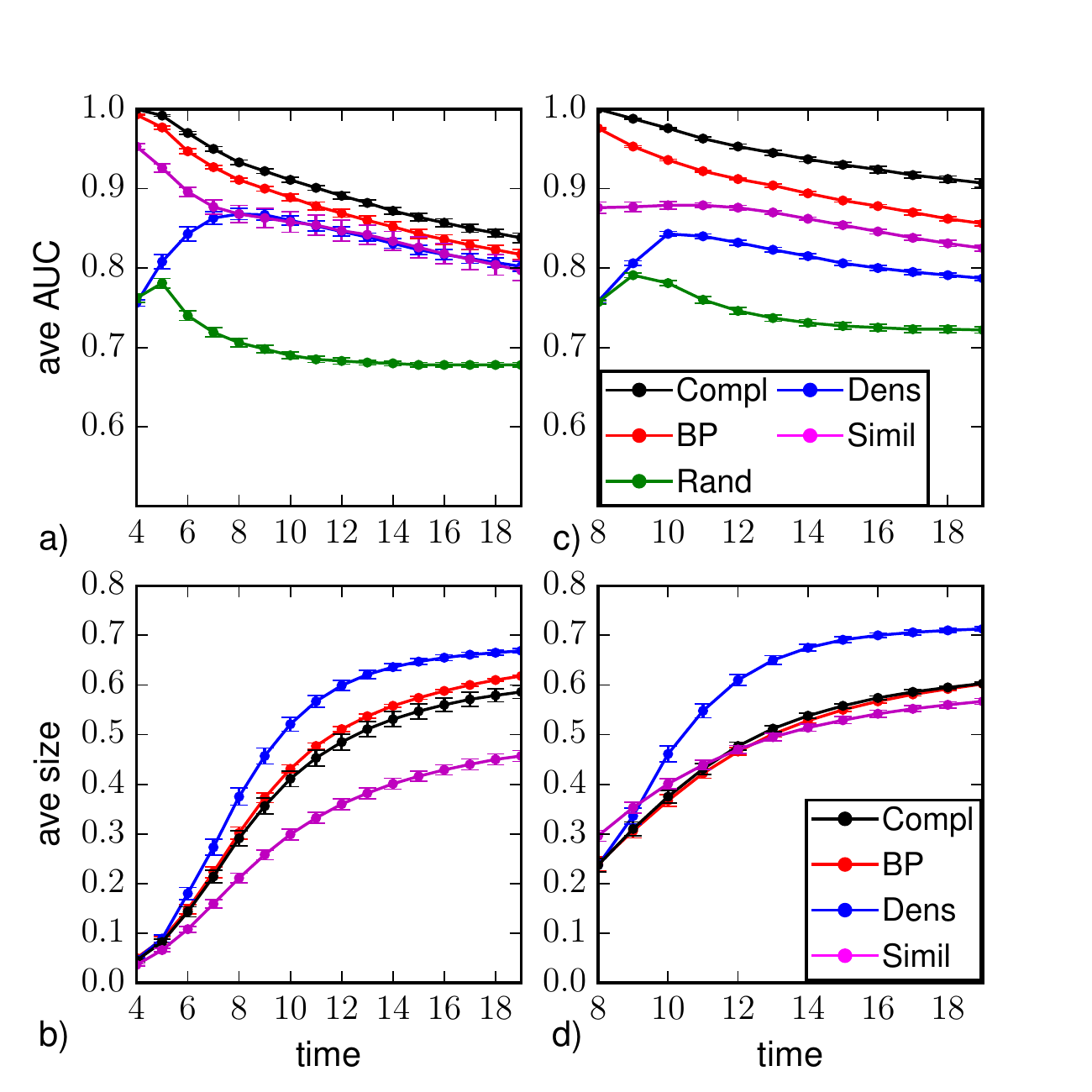}
\caption{Average area under the ROC curve (a,c) and average epidemic size (b,d) as function to the time $t \ge T_{obs}$ for SIR dynamics ($\lambda=0.5$, $\mu=0.4$) on the SC network. Results are obtained with random sampling (green), density sampling (blue), similarity sampling (magenta) and Belief Propagation (red) from a random observation of $30\%$ of the nodes at $T_{obs}=4$ (a,b) and $T_{obs}=8$ (c,d).  For $T_{obs}=8$, only instances with a number of observed infected and recovered nodes $N_{I+R}>18$ is considered ($75\%$ of instances). For $T_{obs}=4$, only instances with observed infected and recovered nodes $N_{I+R}>6$ is considered ($46\%$ of instances). In all plots direct sampling from a complete observation is shown for comparison (black).
}
\label{fig:casetudy2}
\end{figure}

We remark that results are strongly influenced by the number of infected and recovered nodes in the observation. In this respect, in Fig.~\ref{fig:casetudy2}, we repeat all measurements considering observations at $T_{obs}=4$ containing a number of infected and recovered nodes equal to $N_{I+R}\ge6$ (corresponding to the $46\%$ of all instances), and at $T_{obs}=8$ with $N_{I+R}\ge18$ ($75\%$ of all instances). BP performances improve considerably at $T_{obs}=4$, outperforming all other methods in the case of $T_{obs}=8$. These results can be better understood if we consider that the network is characterized by a well connected core surrounded by many low degree nodes. 
When few infected nodes are observed, they typically are low-degree ones and the epidemic process spreads slowly at early time. In this situation similarity sampling is facilitated because the trajectories leading to the observed states are a small set. On the contrary, it is less accurate when many infected nodes are observed or the observation occurs at later times. However results can likely be improved by an higher computational power. Belief Propagation does not choose the seed among the observed infected and recovered nodes only, it computes the probability of being seed for each node of the network, hence it is more accurate than similarity sampling when many infected and recovered nodes are unobserved. 
When the number of nodes reached by the epidemic spreading at the observation time is small, the effect of the existence of short loops in the network is more important and BP is more likely to overestimate the probability of a node to be infected \cite{altarelli2013large}. It is worth noting that we provide to similarity sampling the information about the initial time $t=0\pm\Delta T_0$ of the epidemic spreading, on the contrary we don't provide such an information to BP.

\begin{figure}[tbh]
\centering
\includegraphics[width=0.9\columnwidth]{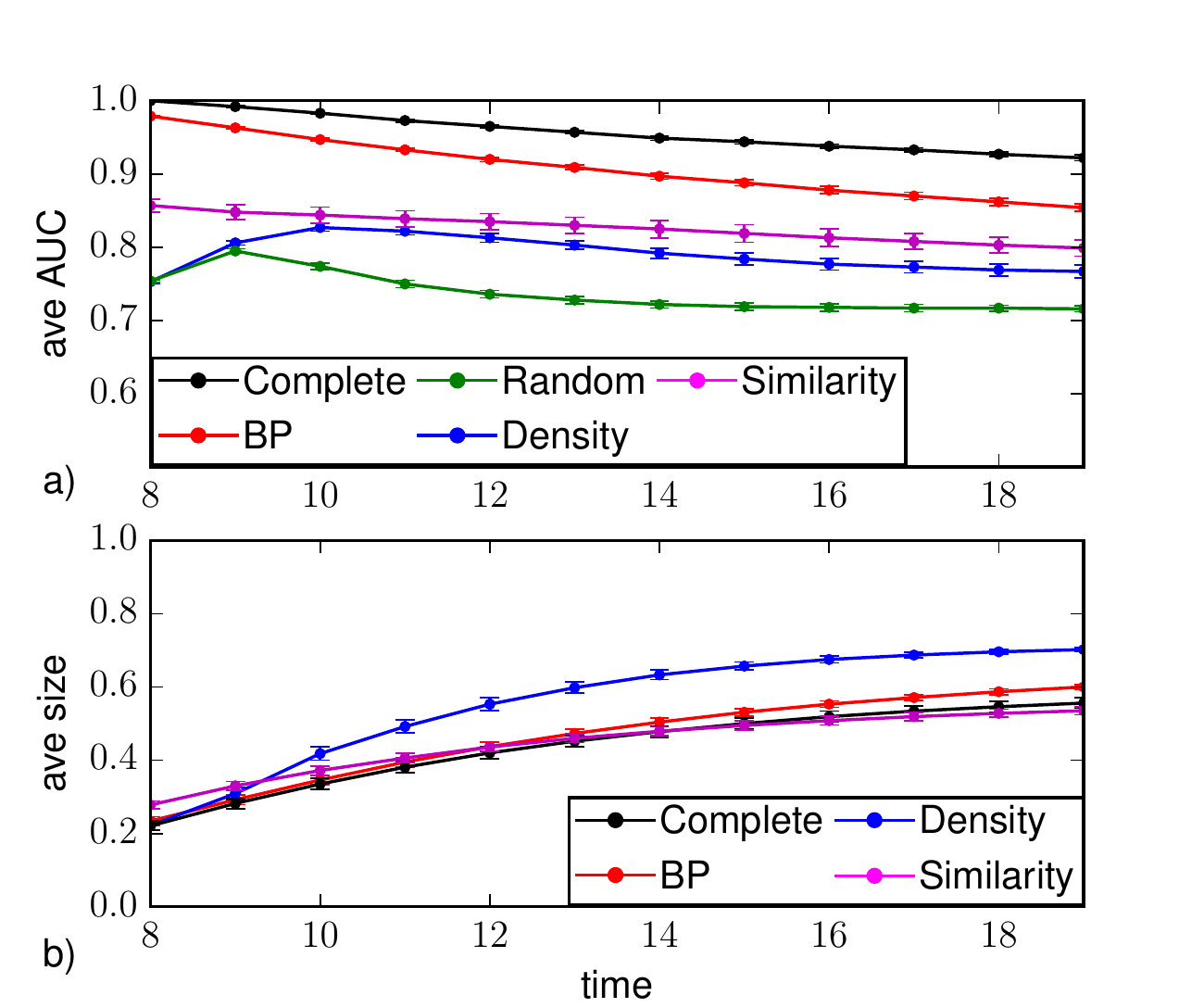}
\caption{Average area under the ROC curve (a) and average epidemic size (b) as function to the time $t \ge T_{obs}$ for SIR dynamics ($\lambda=0.5$, $\mu=0.4$) on the WSC network. Results are obtained with random sampling (green), density sampling (blue), similarity sampling (magenta) and Belief Propagation (red) from a random observation of $30\%$ of the nodes at $T_{obs}=8$. Direct sampling from a complete observation is shown for comparison (black).}
\label{fig:casestudy3}
\end{figure}

We also consider a weighted static projections of the sexual contact network (WSC), in which every existing edge $ij$ is assigned a weight $w_{ij}$ corresponding to the number of contacts between node $i$ and node $j$ during the period under consideration. Then we define the probability that node $i$ infects node $j$ as $\lambda_{ij}=1-(1-\lambda)^{w_{ij}}$. Fig.\ref{fig:casestudy3} shows results for the average AUC and the average epidemic size. Belief Propagation provides higher values for the AUC than all the other methods at all times, even though AUC decreases with time much faster compared to direct sampling with complete observation. Immediately after the observation BP also provided the best prediction of the average epidemic size, while at late times similarity sampling works better.  

\section*{Conclusions}\label{sec:conclusion}
In the present work, we extended the Bayesian Belief Propagation approach to the prediction of the future evolution of an epidemics, providing an efficient distributed algorithm to compute, at any time, the marginal probability of the states of individual nodes in the network. Some global quantities, such as the average epidemic size, can be directly computed as function of the individual marginal probabilities. Here we show that also quantities such as the extinction time distribution, that is intrinsically non local, can be reduced, in the BP approach, to a distributed calculation of local marginals on a locally tree-like factor graph.
On random regular graphs and Barab\'asi-Albert networks, the predictions obtained with the BP algorithm are compared with those from other heuristics based on Monte Carlo direct sampling from the same partial observations, while direct sampling with complete information is taken as a reference to assess the quality of the results. We also analyzed a real-world contact network obtained from a Brazilian database of sexual encounters.
On random networks, BP provides better prediction than the other methods under study at all time steps. 
For all methods, the accuracy of the prediction is lower when the actual number of infected and recovered nodes in the observation is small. The errors introduced in the analysis of these configurations can result in a significant distortion of the overall results, in particular in the long time regime, as observed  in the case of the average epidemic size measured by similarity sampling.
In general, BP is more accurate in the classification of individual marginals than in the estimate of the average epidemic size. A possible reason is that, in some cases, BP equations do not properly converge, resulting in a set of local probability marginals that are slightly different to the correct ones. In particular, convergence issues are mostly due to the presence of small loops in the network, that typically lead to an overestimate of the value of probability marginals by BP. As a consequence, the inaccuracies have little effect on the ranking of individual marginals, on which the ROC classification is based, whereas they are amplified when considering a global quantity such as the average epidemic size.  
Finally, BP usually approximate the extinction time probability distribution better than other methods.

In the real-world case study, results are affected by the presence of a much higher density of edges with respect to random graphs. BP gives better results for observations at late times. When the observation takes place early, the prediction of all methods is clearly worse, but BP gives the best prediction. The inaccurate prediction by BP is probably due to a combined effect of low level of information in the observation and the existence of many short loops in the network that limits the validity of BP. When considering only observations with a sufficiently large number of infected and recovered nodes, BP results improve considerably with respect to all other methods. An evidence of the role played by short loops comes from the better results generally obtained on the weighted network, because by construction the weighted network is effectively more sparse than the unweighted projection first used. 

In conclusion, BP and similarity sampling have advantages and drawbacks depending on the time and type of observation, but BP can be considered the most accurate method to predict both local and global quantities when the underlying network is sparse and when the observation contains a sufficiently large number of infected and recovered nodes. We remark that BP results have been obtained with no knowledge about the initial time of the epidemics, that is another important advantage compared with the other methods.

\section*{Methods}\label{sec:methods}
\subsection*{The SIR epidemic process}
A node $i\in V$ can be in one of the possible states: susceptible (S), infected (I) and recovered (R). At each time step an infected node $i$ can infect each of his neighbor $j$ with a given probability $\lambda_{ij}$, then recover with probability $\mu_i$. The state of a node $i$ at time $t$ is represented by a variable $x_i^t\in\left\{S,I,R\right\}$. The process is irreversible, so once a node recovered it does not get infected anymore. The Markov chain is described by the following transition probabilities
\begin{align}
P(x_i^{t+1}=S|\mathbf{x}^t)&=\mathbb{I}[x_i^{t}=S]\prod_{j\in\partial i}(1-\lambda_{ij}\mathbb{I}[x_i^{t}=I])\\
P(x_i^{t+1}=I|\mathbf{x}^t)&=(1-\mu_i)\mathbb{I}[x_i^{t}=S]+\mathbb{I}[x_i^{t}=S](1-\prod_{j\in\partial i}(1-\lambda_{ij}\mathbb{I}[x_i^{t}=I]))\\
P(x_i^{t+1}=R|\mathbf{x}^t)&=\mathbb{I}[x_i^{t}=R]+\mu_i\mathbb{I}[x_i^{t}=I].
\end{align}
A realization of the SIR stochastic process is univocally expressed in terms of infection times $t_i$ and recovery times $g_i$, $\forall i \in V$. Given the initial configuration $\mathbf{x}^0$, for each node $i\in V$, a recovery time $g_i$ is randomly drawn according to the distribution $\mathcal{G}_i\left(g_i\right)=\mu_i\left(1-\mu_i\right)^{g_i}$ and the infection transmission delays $s_{ij}$ from node $i$ to node $j$ are generated from the conditional distribution 
\begin{equation}
\omega\left(s_{ij}|g_i\right)= \begin{cases}
\lambda_{ij}(1-\lambda_{ij})^{s_{ij}} & s_{ij}\le g_{i}\\
\sum_{s>g_i}\lambda_{ij}(1-\lambda_{ij})^s & s_{ij}=\infty.
\end{cases}
\end{equation}
Infection times are then given by the deterministic equation
\begin{equation}
t_i=\min_{j\in\partial i}(t_j+s_{ji})+1.
\end{equation}

\subsection*{Factor graph representation}
Every realization of the trajectory $({\mathbf x}^0, \dots, \mathbf{x}^t)$ is in one-to-one correspondence with a static configuration of individual infection times $\mathbf{t} =\{t_i\}_{i\in V}$ and recovery times ${\mathbf g} = \{g_i\}_{i \in V}$. Using this static representation of the epidemic dynamics we can express the posterior probability as
\begin{align} 
P\left(\mathbf{x}^{t}|\mathbf{x}^{T_{obs}}\right)&\propto\sum_{\mathbf{t,g,x}^0}P\left(\mathbf{x}^t|\mathbf{t,g}\right)P\left(\mathbf{x}^{T_{obs}}|\mathbf{t,g}\right)P\left(\mathbf{t,g}|\mathbf{x}^0\right)P\left(\mathbf{x}^0\right)
\label{eq:posterior1}
\end{align}
where $P\left(\mathbf{t,g}|\mathbf{x}^0\right)$ is the joint probability distribution of infection and recovery times conditioned on the initial configuration ${\mathbf x}^0$, and  $P\left(\mathbf{x}^t|\mathbf{t,g}\right)$ and $P\left(\mathbf{x}^{T_{obs}}|\mathbf{t,g}\right)$ are deterministic functions of the set $\left(\mathbf{t}, \mathbf{g}\right)$ representing the connection between a $(\mathbf{t}, \mathbf{g})$ configuration and configurations $\mathbf{x}^t$ and $\mathbf{x}^{T_{obs}}$. Although the sum on the right-hand side of Eq.~\eqref{eq:posterior1} still runs over a possibly huge number (exponentially large in $N$) of configurations, a representation in which the dynamical relationships between trajectories of neighboring variables is reduced to a set of local constraints on the activation/recovery times is more convenient to develop approximation methods using tools from graphical models and statistical mechanics. By means of Bayes' theorem we compute the posterior probability of a configuration $\mathbf{x}^t$ at time $t$ given an observation $\mathbf{x}^{T_{obs}}$ at time $T_{obs}$:  
\begin{align}\nonumber
P\left(\mathbf{x}^{t}|\mathbf{x}^{T_obs}\right)&=\sum_{\mathbf{t,g}}{P\left(\mathbf{x}^t|\mathbf{t,g}\right)P\left(\mathbf{t,g}|\mathbf{x}^{T_{obs}}\right)}\\ \nonumber
&\propto\sum_{\mathbf{t,g}}P\left(\mathbf{x}^t|\mathbf{t,g}\right)P\left(\mathbf{x}^{T_{obs}}|\mathbf{t,g}\right)P\left(\mathbf{t,g}\right)=\\ \nonumber
&=\sum_{\mathbf{t,g,x}^0}P\left(\mathbf{x}^t|\mathbf{t,g}\right)P\left(\mathbf{x}^{T_{obs}}|\mathbf{t,g}\right)P\left(\mathbf{t,g}|\mathbf{x}^0\right)P\left(\mathbf{x}^0\right)\\
\label{eq:posterior}
\end{align}
where $P\left(\mathbf{x}^0\right)= \prod_i {\gamma_i(x_i^0)}$ with 
\begin{equation}
\gamma_0(x_i^0)=\gamma\mathbb{I}[x_i^0=I]+(1-\gamma)\mathbb{I}[x_i^0=S]
\end{equation}
is the factorized prior on the initial condition, and $P\left(\mathbf{x}^t|\mathbf{t,g}\right)$ and $P\left(\mathbf{x}^{T_{obs}}|\mathbf{t,g}\right)$ are deterministic functions of the set $\left(\mathbf{t}, \mathbf{g}\right)$ representing the probability of a configuration $\mathbf{x}^t$ at time $t$ (and, respectively, at time $T_{obs}$):

\begin{equation}
P\left(\mathbf{x}^t|\mathbf{t,g}\right)=\prod_i \zeta_i^t\left(t_i,g_i,x_i^t\right)
\end{equation} 
with
\begin{equation}
\zeta_i^t\left(t_i,g_i,x_i^t\right)=\mathbb{I}\left[x_i^t=S, t< t_i\right]+ \mathbb{I}\left[ x_i^t=I, t_i \le t \le (t_i+g_i) \right]+\mathbb{I}\left[ x_i^t=S, t< t_i\right].
\end{equation}

The joint probability distribution of infection and recovery times conditioned on the initial configuration reads

\begin{align}\nonumber
P\left(\mathbf{t,g}|\mathbf{x}^0\right)=& \sum_{s_{ij}}P(\mathbf{s}|\mathbf{g})P(\mathbf{t}|\mathbf{x}^0,\mathbf{g,s}) P(\mathbf{g})\\
=&\sum_{s_{ij}}\prod_{i,j}\omega_{ij}(s_{ij}|g_i)\prod_i\phi_i(t_i,\{t_k,s_{ki}\}_{k\in\partial i}) \mathcal{G}_i(g_i)
\end{align} 
where
\begin{equation}
\phi_i\left(t_i,\left\{t_k,s_{ki}\right\}_{k\in\partial i}\right)=\delta(t_i,\mathcal{I}[x_i^0\ne I](\min_{k\in\partial i}(t_k+s_{ki})+1)).
\end{equation}

The factor graph representation of a probability distribution is made up of a bipartite graph composed of factor nodes and variable nodes\cite{mezard2009information}. Each factorized term in (\ref{eq:posterior}) is represented by a factor node and each variable of the problem is represented by a variable node. Each factor node is connected to the set of variable nodes involved in the corresponding factorized term. The factor graph of (\ref{eq:posterior}) has a loopy structure which can compromise the accuracy of the BP approximation. We can use a factor graph representation that maintains the same topological properties of the original graph in order to guarantee that BP is exact when the underlying graph is a tree. Following  \cite{altarelli2013large,altarelli2013optimizing}, we do that by grouping pairs of variable nodes $\left(t_i, t_j\right)$ in the same variable node. For each edge $(i,j)$ emerging from node $i$ we introduce a triplet $(t_i^{(j)}, t_{ji}, g_i^{j})$, where $t_i^{j}$ and $g_i^{j}$ are copies (on $j$) of the infection time and recovery time of $i$ and the variables $t_{ij}=t_i^{(j)}+s_{ij}$ on which factors $\phi_i$ depend. Including a constraint that forces copies $t_i$ and $g_i$ to have a common value, we get 
\begin{equation}
\psi_i=\delta(t_i, \mathbb{I}[x^0_i\ne I](\min_{j\in\partial i}(t_{ji}+1)))\prod_{j\in\partial i}\delta(t_i^{(j)}, ti)\delta(g_i^{(j)},g_i)
\end{equation}
and
\begin{equation}
\phi_{ij}= \omega_{ij}(t_{ij}-t_i^{(j)}|g_i^{(j)})\omega_{ji}(t_{ji}-t_j^{(i)}|g_j^{(i)}).
\end{equation}
In this representation we can write the posterior probability as the following {\em graphical model}:
\begin{align}
P(\mathbf{x}^t|\mathbf{x}^{T_{obs}})&=\sum_{\mathbf{t,g,s}}\prod_{i,j}\omega_{ij}\prod_i\phi_i\mathcal{G}_i\gamma_i\zeta_i^{T_{obs}}\zeta_i^t.
\end{align}

\subsection*{Belief Propagation Equations}
Given a set $\underline{x}=(x_1,\dots,x_N)$ of random variables with a joint probability distribution
\begin{equation}
M\left(\underline{z}\right) = \frac{1}{Z}\prod_a F_a\left(\underline{z}_{\partial a}\right) 
\end{equation}
where $\underline{z}_{\partial a}\equiv\left\{x_i|i\in\partial a\right\}$ is the set of variables involved in the constraint $a$. Messages are associated with every directed edge on the factor graph and they take values in the space of single-variable probability distributions. The following equations for messages are solved by iteration :

\begin{align*}
p_{F_a\rightarrow i}\left(z_i\right) &= \frac{1}{Z_{ai}}\sum_{\left\{z_j:j\in \partial a\setminus i \right\}}{F_a\left(\left\{z_i\right\}_{i\in\partial a} \right)\prod_{j\in\partial a\setminus i}{m_{j \rightarrow F_a}\left(z_j\right)}}\\
m_{i\rightarrow F_a}\left(z_i\right) &=\frac{1}{Z_{ia}}\prod_{b\in\partial i\setminus a}{p_{F_b\rightarrow i}\left(z_i\right)}\\
m_i\left(z_i\right) &=\frac{1}{Z_i}\prod_{b\in\partial i}{p_{F_b\rightarrow i}\left(z_i\right)}.
\end{align*}
At the fixed point they provide an approximate value for the variables marginal probability \cite{mezard2009information}. In our case the factors $F_a$ are $\psi_i, \phi_{ij}, \gamma_i, \zeta_i^{T_{obs}}, \zeta_i^t$ and $\mathcal{G}_i$ and the variables $z_i$ are couples $(t_i,g_i)$, and triplets $(t_i^{(j)}, g_i^{(j)}, t_{ij})$, $(t_i, g_i, x^{T_{obs}}_i)$, $(t_i, g_i, x^{t}_i)$. The explicit form for the update equations of the $\psi_i$ factor nodes is:
\begin{align}
p_{\psi_i\rightarrow j}(t_i^{(j)},t_{ji},g_i^{(j)})&\propto\sum_{g_i,t_i}\sum_{\left\{t_i^{(k)},t_{ki},g_i^{(k)}\right\}}m_{i\rightarrow\psi_i}(t_i,g_i)\times\\
&\times\prod_{k\in\partial i\setminus j} m_{k\rightarrow\psi_i}(t_i^{(k)},t_{ki},g_i^{(k)})\psi_i\left(t_i,g_i,\left\{(t_i^{(k)},t_{ki},g_i^{(k)})\right\}_{k\in\partial i}\right)
\end{align}
and
\begin{equation}
p_{\psi_i\rightarrow j}(t_i, g_i)\propto\sum_{\left\{t_i^{(k)},t_{ki},g_i^{(k)}\right\}}\prod_{k\in\partial i\setminus j} m_{k\rightarrow\psi_i}(t_i^{(k)},t_{ki},g_i^{(k)})\psi_i\left(t_i,g_i,\left\{(t_i^{(k)},t_{ki},g_i^{(k)})\right\}_{k\in\partial i}\right).
\end{equation}
Efficient forms for these update equations are given in \cite{PhysRevLett.112.118701,1742-5468-2014-10-P10016}.\\
In the factor graph representation, the Bethe free-energy of the graphical model can be expressed as (see also\cite{mezard2009information})
\begin{equation}
-f= \sum_a{f_a}+\sum_i{f_i}-\sum_{(ia)}{f_{(ia)}} 
\end{equation}
in which the local contributions can be expressed as function of the Belief Propagation messages 
\begin{align}
f_a &= \log\left(\sum_{\left\{z_i:i\in\partial a\right\}}{F_a(\left\{z_i\right\}_{i\in\partial a})\prod_{i\in\partial a}m_{i\rightarrow a}(z_i)}\right)\\
f_i &= \log\left(\sum_{z_i}\prod_{b\in\partial i}{p_{F_b\rightarrow i}(z_i)}\right)\\
f_{(ia)} &= \log\left(\sum_{z_i}m_{i\rightarrow a}(z_i)p_{F_b\rightarrow i}(z_i)\right).
\end{align}

\subsection*{The extinction-time constraint}
The posterior probability $P\left(T_{ext}| \mathbf{x}^T_{obs}\right)$ of the extinction time from a partial observation at $T_{obs}$ can be written as
a difference of posterior probabilities that an epidemic ends within a given time,
\begin{equation}
P\left(T_{ext}|\mathbf{x}^T_{obs}\right)=P\left(t_{ext}< T_{ext}|\mathbf{x}^T_{obs}\right)- P\left(t_{ext}< T_{ext}-1|\mathbf{x}^T_{obs}\right).
\end{equation}
Using the static representation of dynamical trajectories,  
\begin{align}
P\left(t_{ext} < T_{ext}| \mathbf{x}^T_{obs}\right) &= \sum_{\mathbf{t,g}}{P\left(t_{ext} < T_{ext}|\mathbf{t,g}\right)P\left(\mathbf{t,g}|\mathbf{x}^{T_{obs}}\right)}\\
&\propto\sum_{\mathbf{t,g}}P\left(t_{ext}<T_{ext}|\mathbf{t,g}\right)P\left(\mathbf{x}^{T_{obs}}|\mathbf{t,g}\right)P\left(\mathbf{t,g}\right)\\
&=\sum_{\mathbf{t,g,x}^0}P\left(t_{ext}<T_{ext}|\mathbf{t,g}\right)P\left(\mathbf{x}^{T_{obs}}|\mathbf{t,g}\right)P\left(\mathbf{t,g}|\mathbf{x}^0\right)P\left(\mathbf{x}^0\right)\\
&=\sum_{\mathbf{t,g,x^0}}{\mathcal{Q}\left(\mathbf{x}^{T_{obs}}, T_{ext},\mathbf{t}, \mathbf{g},x^0\right)}= Z\left(T_{ext}, \mathbf{x}^{T_{obs}}\right).
\end{align}

The terms in the latter expression are the same as in \eqref{eq:posterior}, with the exception of the following term factorized over the nodes
\begin{equation}
P\left(t_{ext}<T_{ext}|\mathbf{t,g}\right)= \prod_i{\mathbb{I}\left[\left(t_i+g_i\right)<T_{ext}\right]}
\end{equation}
that constrains the calculation to epidemics that vanish before $T_{ext}$. In this way, we give null probability to every single site configuration with $t_i+g_i$ larger than $T_{ext}$ (except for $t_i=T_{inf}$ that describes susceptible nodes). The logarithm of the partition function is the free energy of the model, hence
\begin{equation}
-f\left(T_{ext}, \mathbf{x}^T_{obs}\right) = \log{Z\left(T_{ext}, \mathbf{x}^T_{obs}\right)}= \log P\left(t_{ext}< T_{ext}| \mathbf{x}^T_{obs}\right).
\end{equation} 
In the factor graph representation, the free-energy can be approximated with the Bethe free-energy, that is computed by means of the BP equations.

\section*{Acknowledgements}
LDA acknowledges the Italian FIRB Project No. RBFR10QUW4 and ERC grant No. 267915. 
AB and LDA acknowledge support by Fondazione CRT under the initiative  ``La Ricerca dei Talenti''.

\bibliography{biblio}{}

\begin{thebibliography}{10}

\bibitem{1742-5468-2014-10-P10016}
F.~Altarelli, A.~Braunstein, L.~Dall'Asta, A.~Ingrosso, and R.~Zecchina.
\newblock The patient-zero problem with noisy observations.
\newblock {\em Journal of Statistical Mechanics: Theory and Experiment},
  2014(10):P10016, 2014.

\bibitem{PhysRevLett.112.118701}
F.~Altarelli, A.~Braunstein, L.~Dall'Asta, A.~Lage-Castellanos, and
  R.~Zecchina.
\newblock Bayesian inference of epidemics on networks via belief propagation.
\newblock {\em Phys. Rev. Lett.}, 112:118701, Mar 2014.

\bibitem{PhysRevX.4.021024}
F.~Altarelli, A.~Braunstein, L.~Dall'Asta, J.~R. Wakeling, and R.~Zecchina.
\newblock Containing epidemic outbreaks by message-passing techniques.
\newblock {\em Phys. Rev. X}, 4:021024, May 2014.

\bibitem{altarelli2013large}
F.~Altarelli, A.~Braunstein, L.~Dall'Asta, and R.~Zecchina.
\newblock Large deviations of cascade processes on graphs.
\newblock {\em Physical Review E}, 87(6):062115, 2013.

\bibitem{altarelli2013optimizing}
F.~Altarelli, A.~Braunstein, L.~Dall'Asta, and R.~Zecchina.
\newblock Optimizing spread dynamics on graphs by message passing.
\newblock {\em Journal of Statistical Mechanics: Theory and Experiment},
  2013(09):P09011, 2013.

\bibitem{antulov2015identification}
N.~Antulov-Fantulin, A.~Lan{\v{c}}i{\'c}, T.~{\v{S}}muc,
  H.~{\v{S}}tefan{\v{c}}i{\'c}, and M.~{\v{S}}iki{\'c}.
\newblock Identification of patient zero in static and temporal networks:
  Robustness and limitations.
\newblock {\em Physical review letters}, 114(24):248701, 2015.

\bibitem{CLM:CLM12472}
A.~Barrat, C.~Cattuto, A.~E. Tozzi, P.~Vanhems, and N.~Voirin.
\newblock Measuring contact patterns with wearable sensors: methods, data
  characteristics and applications to data-driven simulations of infectious
  diseases.
\newblock {\em Clinical Microbiology and Infection}, 20(1):10--16, 2014.

\bibitem{BelikPRX11}
V.~Belik, T.~Geisel, and D.~Brockmann.
\newblock Natural human mobility patterns and spatial spread of infectious
  diseases.
\newblock {\em Phys. Rev. X}, 1:011001, Aug 2011.

\bibitem{Braunstein2016inference}
A.~Braunstein and A.~Ingrosso.
\newblock Inference of causality in epidemics on temporal contact networks.
\newblock {\em Scientific reports}, 6:27538, 2016.

\bibitem{cattuto2010dynamics}
C.~Cattuto, W.~Van~den Broeck, A.~Barrat, V.~Colizza, J.-F. Pinton, and
  A.~Vespignani.
\newblock Dynamics of person-to-person interactions from distributed rfid
  sensor networks.
\newblock {\em PloS one}, 5(7):e11596, 2010.

\bibitem{colizza2006modeling}
V.~Colizza, A.~Barrat, M.~Barth{\'e}lemy, and A.~Vespignani.
\newblock The modeling of global epidemics: Stochastic dynamics and
  predictability.
\newblock {\em Bulletin of mathematical biology}, 68(8):1893--1921, 2006.

\bibitem{colizza2007predictability}
V.~Colizza, A.~Barrat, M.~Barth{\'e}lemy, and A.~Vespignani.
\newblock Predictability and epidemic pathways in global outbreaks of
  infectious diseases: the sars case study.
\newblock {\em BMC medicine}, 5(1):34, 2007.

\bibitem{holme2013extinction}
P.~Holme.
\newblock Extinction times of epidemic outbreaks in networks.
\newblock {\em PloS one}, 8(12), 2013.

\bibitem{holme2015information}
P.~Holme.
\newblock Information content of contact-pattern representations and
  predictability of epidemic outbreaks.
\newblock {\em arXiv preprint arXiv:1503.06583}, 2015.

\bibitem{Holme201297}
P.~Holme and J.~Saramäki.
\newblock Temporal networks.
\newblock {\em Physics Reports}, 519(3):97 -- 125, 2012.
\newblock Temporal Networks.

\bibitem{holme2015time}
P.~Holme and T.~Takaguchi.
\newblock Time evolution of predictability of epidemics on networks.
\newblock {\em Physical Review E}, 91(4):042811, 2015.

\bibitem{mastrandrea2015contact}
R.~Mastrandrea, J.~Fournet, and A.~Barrat.
\newblock Contact patterns in a high school: a comparison between data
  collected using wearable sensors, contact diaries and friendship surveys.
\newblock {\em PloS one}, 10(9):e0136497, 2015.

\bibitem{mezard2009information}
M.~Mezard and A.~Montanari.
\newblock {\em Information, physics, and computation}.
\newblock Oxford University Press, 2009.

\bibitem{pastor2015epidemic}
R.~Pastor-Satorras, C.~Castellano, P.~Van~Mieghem, and A.~Vespignani.
\newblock Epidemic processes in complex networks.
\newblock {\em Reviews of modern physics}, 87(3):925, 2015.

\bibitem{rocha2010information}
L.~E. Rocha, F.~Liljeros, and P.~Holme.
\newblock Information dynamics shape the sexual networks of internet-mediated
  prostitution.
\newblock {\em Proceedings of the National Academy of Sciences},
  107(13):5706--5711, 2010.

\bibitem{rocha2011simulated}
L.~E. Rocha, F.~Liljeros, and P.~Holme.
\newblock Simulated epidemics in an empirical spatiotemporal network of 50,185
  sexual contacts.
\newblock {\em PLoS Comput Biol}, 7(3):e1001109, 2011.

\bibitem{salathe2010high}
M.~Salath{\'e}, M.~Kazandjieva, J.~W. Lee, P.~Levis, M.~W. Feldman, and J.~H.
  Jones.
\newblock A high-resolution human contact network for infectious disease
  transmission.
\newblock {\em Proceedings of the National Academy of Sciences},
  107(51):22020--22025, 2010.

\bibitem{stehlePONE11}
J.~Stehlé, N.~Voirin, A.~Barrat, C.~Cattuto, L.~Isella, J.~Pinton,
  M.~Quaggiotto, W.~{Van den Broeck}, C.~Régis, B.~Lina, and P.~Vanhems.
\newblock High-resolution measurements of face-to-face contact patterns in a
  primary school.
\newblock {\em PLOS ONE}, 6(8):e23176, 08 2011.

\end{thebibliography}

\end{document}